\documentclass[prd,preprintnumbers,showpacs,onecolumn,showkeys,floatfix,nofootinbib,notitlepage]{revtex4-1}
\usepackage{graphicx}
\usepackage{amssymb}
\usepackage{amsmath}
\usepackage{epsfig}
\usepackage{color}
\usepackage{enumerate}
\usepackage{slashed}
\usepackage{hyperref}
\usepackage{epstopdf}
\usepackage{axodraw4j}
\usepackage{pstricks}

\def\slashchar#1{\setbox0=\hbox{$#1$}
   \dimen0=\wd0 \setbox1=\hbox{/} \dimen1=\wd1
   \ifdim\dimen0\big>\dimen1 \rlap{\hbox to \dimen0{\hfil/\hfil}} #1
   \else  \rlap{\hbox to \dimen1{\hfil$#1$\hfil}} / \fi}

\newcommand{\ud}{\mathrm{d}}
\newcommand{\be}{\begin{equation}}
\newcommand{\ee}{\end{equation}}
\newcommand{\bea}{\begin{eqnarray}}
\newcommand{\eea}{\end{eqnarray}}

\newcommand{\Appendix}[1]%
    {%
     \section{#1}%
      }

\newcounter{RomanNumber}
\newcommand{\Rum}[1]
{\setcounter{RomanNumber}
{#1}\Roman{RomanNumber}}

\begin{document}

\title{Glauber Gluon Effects in Soft Collinear Factorization}

\author{Gao-Liang Zhou$^{1}$}
\email{ zhougl@alumni.itp.ac.cn}
\author{Zheng-Xin Yan$^{1}$, Xin Zhang$^{1}$ and Feng Li$^{2}$}
\affiliation{$^{1}$College of Science, Xi'an University of Science and Technology, Xi'an 710054, People's Republic of China}
\affiliation{$^{2}$College of Science,   Nanjing University of Posts and Telecommunications , Nanjing 210023, People's Republic of China}




\begin{abstract}
Effects of Glauber gluons, which cause the elastic scattering process between different jets, are studied in the frame of soft-collinear effective theory(SCET). Glauber modes are added into the Lagrangian before integrated out, which is helpful in studies on Glauber couplings of collinear and soft particles explicitly. It is proved that interactions after the hard collision cancel out in processes inclusive enough. So are interactions with the light cone coordinates $x^{+}$ and $x^{-}$ greater than those of the hard collision. Eikonalization of Glauber couplings of active particles and absorption of active-active and active-soft and active-spectator Glauber exchanges into soft and collinear Wilson lines are discussed, which is related to loop level definitions of Glauber gluons here.    The active-spectator coherence are proved to be harmless on inclusive summation of spectator-spectator and spectator-soft  Glauber exchanges. Based on this result, spectator-spectator and spectator-soft Glauber exchanges are proved to cancel out in processes considered here.  Graphic aspects of the cancellation are also discussed to explain relations between the graphic cancellation of Glauber gluons in the frame of perturbative QCD and operator level skills in the paper.
\end{abstract}


\maketitle

\section{Introduction.}
\label{introduction}

Soft collinear factorization is crucial in connecting perturbative QCD calculations with experimental data of high energy hadron processes.  Such factorization is proved in many typical process in the frame of perturbative QCD,(see, for example, Refs.\cite{CS:1982,B:1985,CSS:1985,CSS:1988,CSS:1989,S:1993,Collins:2013,Diehl:2015bca}). Comparing to these results, factorization in soft-collinear effective theory(SCET)\cite{SCET:2001,SCET:2002,BFPRS:2002,Wei:2003ns} provides us new perspectives to understand QCD factorization, which seems more intuitive. SCET describe interactions among  collinear, soft and ultrasoft particles. Especially, hard collisions between collinear particles are described by various effective operators in SCET. In original SCET Lagrangian, ultrasoft gluons decouple from collinear fields after an unitary transformation. Effects of couplings between collinear particles and soft gluons are absorbed into soft Wilson lines. Thus  soft collinear factorization holds at Lagrangian level in original SCET.

Despite of great advantages, the original SCET did not deal with Glauber gluons properly. Glauber gluons, which take space like momenta,  are responsible for  elastic scattering processes between different jets.  Results in \cite{BBL:1981} display how Glauber gluons break the QCD factorization in processes involving two initial hadrons like the Drell-Yan process. For Drell-Yan process,  leading pinch singularities in Glauber region cancel out  according to unitarity\cite{B:1985,CSS:1985,CSS:1988}.  One may then deform the integral path of loop momenta to avoid the Glauber region. After the deformation, couplings between Glauber gluons and collinear particles eikonalize. That is, Glauber gluons behave like soft or collinear gluons after the deformation. In summary, soft collinear factorization of Drell-Yan process is not violated by Glauber gluons\cite{B:1985,CSS:1985,CSS:1988}.   However, the factorization can be violated by Glauber gluons for processes which are not inclusive enough(see, for example, Refs.\cite{CFS:1993,CDF:1997,Alvero:1997rw,Goulianos:1997pp,H1:2007}). In these processes, cancellation of leading pinch singularities in Glauber region is hindered by Glauber couplings of  detected final particles

Although leading pinch singular singularities in Glauber region may cancel out,  Glauber gluon effects are visible in processes in which the factorization works. Since the contour deformation to avoid Glauber region relies on explicit processes, wether the collinear and soft Wilson lines appearing in parton(distribution and fragmentation) functions and soft factors are past-pointing  or future-pointing is process dependent\cite{CM:2004}. In other words, Glauber gluons affect directions of various Wilson lines in factorization. Such effects are more obvious in transverse momentum dependent(TMD) objects like the Sivers function\cite{Sivers:1989cc,Collions:2002,Kang:2009bp},  although these objects are not concerned here.

Glauber gluons in SCET are more subtle. In \cite{Idilbi:2008vm,DEramo:2010wup,Ovanesyan:2011xy,Benzke:2012sz,Ovanesyan:2012fr}, Glauber gluon fields are added into the SCET Lagrangian to describe jets in dense QCD matter.  The effective theory is termed as $\text{SCET}_{\text{G}}$. Glauber gluons in $\text{SCET}_{\text{G}}$ behaves like QCD background and does not cause scattering between different jets directly. This is different form the situation one confronts in usual soft collinear factorization. In \cite{Bauer:2010cc,Fleming:2014rea,Rothstein:2016bsq}, Glauber gluons are integrated out and effective operators that describe elastic scattering effects between collinear particles are introduced into the effective action of SCET. These operators are nonlocal, in which the decoupling of ultrasoft gluons from collinear fields are no longer manifest.  Matching of coefficients of these operators relies on suitable subtraction formalism\cite{Bauer:2010cc,Manohar:2006nz} to avoid double counting in loop integrals and systematic scheme\cite{Chiu:2011qc,Chiu:2012ir} to regularize rapidity divergences.
These effective operators may violate the factorization theorem in SCET as they cause coherence between different jets. In \cite{Rothstein:2016bsq}, authors discuss properties of these operators. Especially,  cancellation  of spectator-spectator type Glauber exchanges and absorptions of spectator-active and  active-active type Glauber exchanges into collinear and soft Wilson lines are discussed in \cite{Rothstein:2016bsq}. These discussions are necessary for proofs of factorization theorem in SCET. However, they are not enough as these discussions are restricted to  ladder like diagrams. More discussions on these topics are necessary for  proofs of factorization theorem in SCET. This is the primary motivation of our paper.

Glauber gluons should be viewed as modes different from collinear and soft gluons no matter wether they are integrated out or not. The question is how to distinguish them from collinear and soft gluons in loop integral? In other words, how to avoid double counting of contributions of Glauber gluons and other modes in loop integral? Let us start from approximations to describe couplings between Glauber gluons and other particles. For example, let us consider  a Glauber gluon $q$ exchanged between plus- and minus-collinear particles. At leading infrared power, one has
\begin{equation}
\label{Glauberapp}
\frac{-i}{l^{2}+i\epsilon}\simeq \frac{-i}{(l_{\perp})^{2}+i\epsilon}.
\end{equation}
One can also neglect $l^{+}$($l^{-}$) in couplings between $l$ and plus(minus)-collinear particles.
These approximations are helpful to describe Glauber couplings in loop integral. Let us consider the coupling between a gluon $q$ and a plus-collinear particle $k$ further. If $q$ is soft or minus-collinear, then the coupling between $q$ and $k$ eikonalize. The eikonalized part of the coupling can be absorbed into collinear or soft Wilson lines and  dose not affect the factorization even if $q$ locates in Glauber region. The non-eikonalized part should be power suppressed in minus-collinear or soft(ultrasoft) region of $q$. Hence on has
\begin{equation}
|q^{+}|\gg |q_{\perp}|,\quad |q^{+}q^{-}|\simeq |q_{\perp}|^{2}
\end{equation}
or
\begin{equation}
|q^{+}q^{-}|\ll |q_{\perp}|^{2}
\end{equation}
at leading infrared power. In other words, $q$ should be plus-collinear or Glauber for the non-eikonalized part. If $q$ is plus-collinear, then couplings between $q$ and minus-collinear and soft particles should eiknalize.\footnote{Coulings between Glauber gluons and those between Glauber gluons and ultrsoft particles are power suppressed as discussed in Sec.\ref{SCETg}}  After subtraction of collinear and soft and ultrasoft region of $q$, one can take the approximation (\ref{Glauberapp}) for $q$. This is how one should define Glauber gluons in loop integrals.\footnote{According to this definition, non-Glauber region contributes to loop integrals of Glauber gluon, although their contributions are power suppressed.} Rapidity divergences in this definition can be  controlled through the regulator presented in \cite{Rothstein:2016bsq,Chiu:2011qc,Chiu:2012ir}, which reads
\begin{equation}
\label{regulator}
\omega^{2}|2q^{z}|^{-\eta}\nu^{\eta}\propto |q^{+}-q^{-}|^{-\eta}.
\end{equation}

We find it convenient to introduce Glauber gluon fields into SCET action before integrating them out. This is helpful to determine power counting for couplings involving Glauber gluons.  Especially,  it helps us to see origins of leading power effective operators in \cite{Rothstein:2016bsq}. We should mention that power counting for various modes may depend on explicit gauge conditions in perturbative calculations as shown in\cite{Bauer:2010cc,Fleming:2014rea,Rothstein:2016bsq}.  Compared to the covariant gauge in \cite{Bauer:2010cc,Fleming:2014rea,Rothstein:2016bsq}, it is more convenient to work in the Feynman gauge for issues considered here. There are super leading powers in practical diagrams in the Feynman gauge. However, such super leading powers cancel out in physical observable according to the Ward identity\cite{S:1993,Collins:2013}. This is confirmed by the  power counting result presented in this paper. Thus super leading powers in the Feynman gauge do not disturb us.

\begin{figure*}
\begin{tabular}{cc}
\includegraphics[scale=0.5]{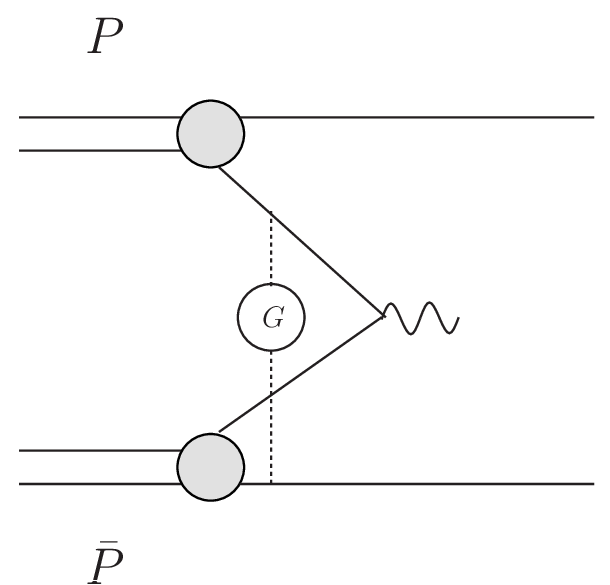}
&
\includegraphics[scale=0.5]{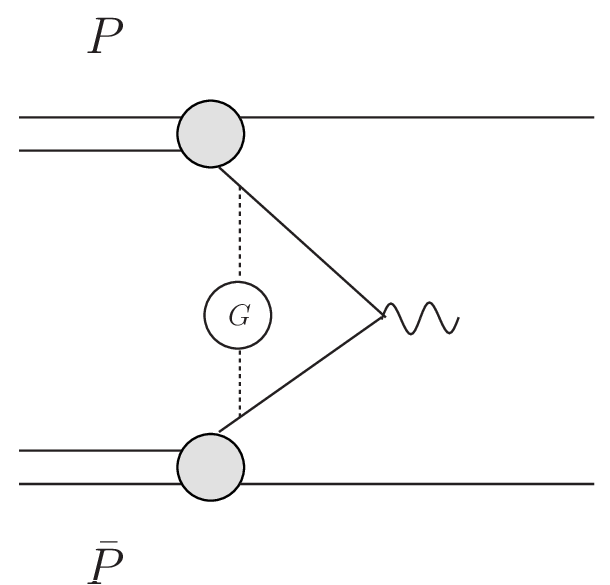}
\\
(a)&(b)
\end{tabular}
\caption{Examples of ladder diagrams with Glauber gluons exchanged (a)between active particles and spectators;(b)between spectators.}
\label{ladder}
\end{figure*}

\begin{figure*}
\begin{center}
\includegraphics[scale=0.5]{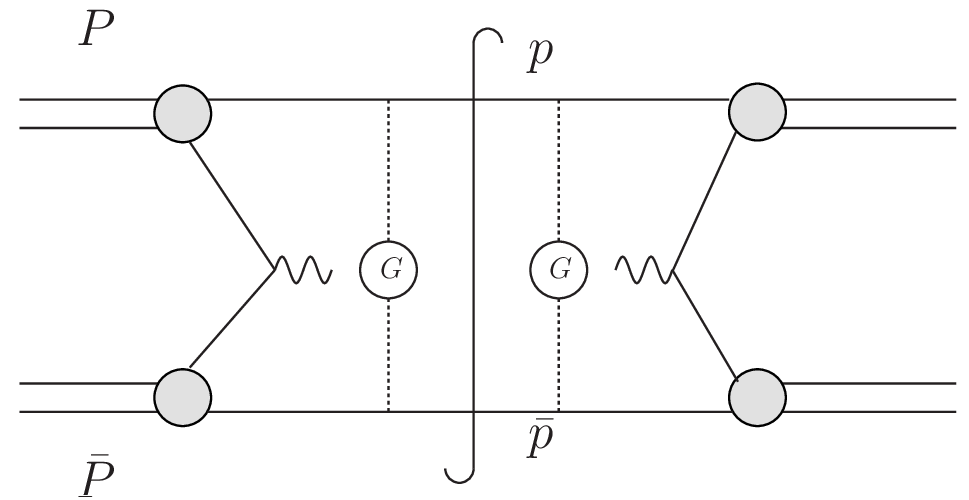}
\\
\includegraphics[scale=0.5]{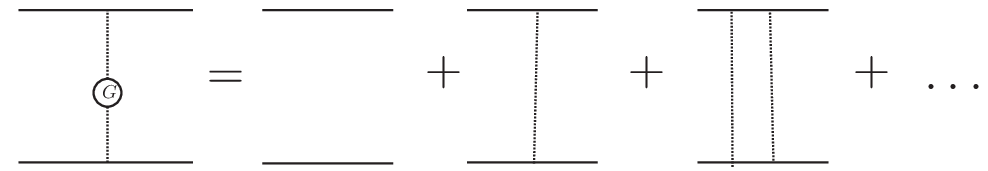}
\end{center}
\caption{Elastic scattering diagrams between spectators, where dot lines represent Glauber gluons.}
\label{ss}
\end{figure*}

Subtraction of eikonalized couplings from definition of Glauber gluon modes is important in treatment of elastic scattering processes between collinear particles. Eikonalized part of couplings involving collinear particles  should be viewed as soft and collinear and ultrasoft interactions of the collinear particles even if there are gluons with Glauber momenta. Especially,  couplings between collinear particles and exchanged gluons eikonalize in active-spectator and active-active  exchanges even if the exchanged gluons take Glauber momenta. Hence these exchanged gluons should be absorbed into dedition of collinear or soft gluons. This is confirmed by results in  \cite{Rothstein:2016bsq}, in which  ladder diagrams like those shown in Fig.(\ref{ladder}) were discussed.
After the approximation (\ref{Glauberapp}), the subtraction dose not affect spectator-spectator type Glauber gluons in dimensional regularization scheme as dimensionless integral vanished in the scheme.   Ladder diagrams of spectator-spectator Glauber exchanges as shown in Fig.\ref{ss} are discussed in \cite{Rothstein:2016bsq}.
According to calculations in \cite{Rothstein:2016bsq},  these ladder diagrams cancel out for processes inclusive enough. Particles exchanged between spectators and active particles are absent in Fig.\ref{ss}. Thus coherence between spectators and active particles is trivial in these discussions. In this paper, we would like to prove that cancellation of spectator-spectator Glauber exchanges is not affected by spectator-active coherence for inclusive processes.

The paper is organized as follows. In Sec.\ref{SCETg}, we add Glauber gluons  into SCET action. Power counting for couplings involving Glauber gluons  is also presented in this section.  In Sec.\ref{sspectator}, we prove the cancellation of interactions after the hard collision for processes inclusive enough.     In Sec.\ref{AG}, we consider couplings between active particles and Glauber gluons. We prove the eikonalization of these couplings and explain why these couplings are equivalent to zero bins of soft and collinear couplings of active particles.   In Sec.\ref{canSS}, we prove the cancellation of spectator-spectator Glauber exchanges for processes inclusive enough. We exclude the influence of  spectator-active coherence by proving that the coherence should occur before spectator-spectator Glauber exchanges. Hence the summation over final spectators without affecting spectator-active coherence is enough for cancellation of spectator-spectator Glauber exchanges.     In Sec.\ref{cangrpah}, we explain how our operator skills are related to graphic cancellation of spectator-spectator Glauber exchanges in \cite{B:1985,CSS:1985,CSS:1988}.  Our conclusions and some discussions are presented in Sec.\ref{conc}.

\section{Glauber Gluons in SCET}
\label{SCETg}

In this section, we introduce Glauber gluon fields into SCET action and study interactions between Glauber gluons and other particles. These discussions are helpful on studies of Glauber effects in hadronic processes at leading power.

Glauber gluons take space like momenta and cause elastic scattering between collinear particles. For example, we consider a gluon with momentum scales as,
\begin{equation}
(p^{+},p^{-},p_{\perp})\sim Q(\lambda^{2},\lambda^{2},\lambda),
\end{equation}
where $Q$ refreshments a hard energy scale and $\lambda\ll 1$. The gluon is space like as $p^{2}\sim -Q^{2}\lambda^{2}<0$. Exchanging of such gluon causes elastic scattering between particles collinear to plus and minus direction. We discuss power counting for couplings between Glauber gluons and other particles in this section. Results in this section are compatible with those in \cite{Rothstein:2016bsq} and make our discussions in our following sections more clear.

For propagation of particles collinear to $n^{\mu}=\frac{1}{\sqrt{2}}(1,\vec{n})$, which is light like, the modes related to the problem and power counting for them in the Feyanman gauge are presented in Table \ref{Power counting}\cite{SCET:2001,SCET:2002,Rothstein:2016bsq}, where $\bar{n}^{\mu}=\frac{1}{\sqrt{2}}(1,-\vec{n})$ and $n\cdot p_{n\perp}=\bar{n}\cdot p=n\cdot A_{n,pn\perp}=\bar{n}\cdot A_{n,pn\perp}=0$.
\footnote{In \cite{SCET:2001,SCET:2002}, authors work in covariant gauge,
\begin{equation}
\int\ud^{4}x e^{ik\cdot x}T<A_{nG}^{\mu}(x)A_{nG}^{\nu}(0)>=\frac{-i}{k^{2}}(g^{\mu\nu}-(1-\xi)\frac{k^{\mu}k^{\nu}}{k^{2}}),
 \end{equation}
 where $\xi$ is the gauge parameter. Power counting for the field $A_{n}$ reads $(n\cdot A_{n,p},\bar{n } \cdot A_{n,p},A_{n,pn\perp})\sim (\lambda^{2},1,\lambda)$. It is required that $1-\xi$ is not too small to get this result. While working in the Feynman gauge, there are super leading power terms involving $n\cdot A_{n,p}$ in SCET Lagrangian. This does not disturb us as $n\cdot A_{n,p}$ and $\bar{n}\cdot A_{n,p}$ appear in pairs in practical diagrams.}

To obtain the power counting for Glauber gluon fields, we consider the propagator of Glauber gluon in the Feynman gauge,
\begin{equation}
\int\ud^{4}x e^{ik\cdot x}T<A_{nG}^{\mu}(x)A_{nG}^{\nu}(0)>=\frac{-ig^{\mu\nu}}{k^{2}}.
 \end{equation}
Momentum of the Glauber gluon scale as $(n\cdot k, \bar{n}\cdot k,k_{n\perp})\sim Q(\lambda^{2},\lambda^{b},\lambda)$ and the integral volume scale as $\int \ud^{4}x\sim Q^{-4}\lambda^{-4-b}$. Thus power counting for Glauber gluon fields reads $A_{nG}^{\mu}\sim \lambda^{1+\frac{b}{2}} $ given that $\xi $ is not too large. For future covalence,  we present here the power counting for fermions with momenta scales as $(n\cdot k, \bar{n}\cdot k,k_{n\perp})\sim Q(\lambda^{2},\lambda^{b},\lambda)$, which reads $\lambda^{\frac{3+b}{2}}$.
\begin{table}
\begin{tabular}{cccc}
\hline
modes & fields &momenta scales $(n\cdot p, \bar{n}\cdot p,p_{n\perp})$ & infrared power counting\\
\hline
Collinear quarks & $\xi_{n,p}$& $Q(\lambda^{2}, 1,\lambda)$ & $\lambda$\\
\hline
Collinear gluons & $A_{n,p}$ & $Q(\lambda^{2}, 1,\lambda)$  & $\lambda$\\
\hline
Soft quarks & $q_{s,p}$  & $Q(\lambda, \lambda,\lambda)$ & $\lambda^{3/2}$\\
\hline
Soft gluons & $A_{s,p}^{\mu}$ & $Q(\lambda, \lambda,\lambda)$ & $\lambda$\\
\hline
Ultrasoft quarks & $q_{us}$ & $Q(\lambda^{2}, \lambda^{2},\lambda^{2})$ & $\lambda^{3}$\\
\hline
Ultrasoft gluons & $A_{us}^{\mu}$ & $Q(\lambda^{2}, \lambda^{2},\lambda^{2})$ & $\lambda^{2}$\\
\hline
Glauber gluons & $A_{nG}^{\mu}$ & $Q(\lambda^{2}, \lambda^{b},\lambda)($b=1,2$)$ & $\lambda^{1+\frac{b}{2}}$\\
\hline
\end{tabular}
\caption{Relevant modes in propagation of collinear particles collinear to $n^{\mu}=\frac{1}{\sqrt{2}}(1,\vec{n})$ and power counting for them in the Feyanman gauge, where $\bar{n}^{\mu}=\frac{1}{\sqrt{2}}(1,-\vec{n})$ and $n\cdot p_{n\perp}=\bar{n}\cdot p=n\cdot A_{n,pn\perp}=\bar{n}\cdot A_{n,pn\perp}=0$.}
\label{Power counting}
\end{table}

\subsection{Power counting for couplings between Glauber gluons and other particles}
\label{powerG}

In this subsection, we consider interactions between Glauber gluons and other particles. We first consider couplings between Glauber gluons and ultrasoft gluons. According to table \ref{Power counting}, power counting for ultrasoft gluon and Glauber gluon reads $\lambda^{3}$, $\lambda^{2}$ and $\lambda^{1+\frac{b}{2}}$. Integral volumes of these couplings scale as $\int \ud^{4}x\sim Q^{-4}\lambda^{-4-b}$. In couplings between Glauber gluons and ultrasoft gluons, there are at least two Glauber gluons and one ultrsoft gluon. Infrared power counting for combination of these fields reads $\lambda^{4+b}$. There is an additional gluon field or momentum operator in these couplings according to Lorentz invariance. Infrared power counting for the additional gluon field reads $\lambda^{1+\frac{b}{2}}$ or $\lambda^{2}$. That of the momentum operator reads $\lambda$ or $\lambda^{2}$. Thus infrared power counting for these couplings reads $\lambda^{r}$, where
\begin{equation}
r\ge 1+4+b+(-4-b)=1.
\end{equation}
That is to say, infrared power counting for these couplings reads $\lambda$ or higher.

For couplings between Glauber gluons and ultrasoft fermions, the situation is similar. Integral volumes of these couplings scale as $\int \ud^{4}x\sim Q^{-4}\lambda^{-4-b}$. In these couplings there are at least one ultrasoft fermion, one Glauber gluon and one fermion with momentum scales as $(n\cdot k, \bar{n}\cdot k,k_{n\perp})\sim Q(\lambda^{2},\lambda^{b},\lambda)$. According to Table \ref{Power counting} and previous texts, infrared power counting for these fields reads $\lambda^{3}$, $\lambda^{1+\frac{b}{2}}$ and $\lambda^{\frac{3+b}{2}}$ respectively. Momenta of these fields are quite small and there are not energy scale which may produce minus power of $\lambda$. Thus infrared power counting for these couplings reads $\lambda^{\frac{3}{2}}$ or higher.

For couplings between Glauber gluons without other type particles, the integral volume scales as $\int \ud^{4}x\sim Q^{-4}\lambda^{-4-b}$. There are not least three Glauber gluons in these couplings. Infrared power counting for combination of these fields reads $\lambda^{3+\frac{3b}{2}}$. There is an additional gluon field or momentum operator in these couplings according to Lorentz invariance. Power clouting for the gluon field reads $\lambda^{1+\frac{b}{2}}$. That of the momentum operator reads $\lambda$ or $\lambda^{2}$. Thus power counting for these couplings reads $\lambda^{\frac{b}{2}}$ or higher.

For couplings between Glauber gluon fields $A_{nG}$ and $A_{\bar{n}G}$, which involves soft gluons, the integral volume scales as $\int \ud^{4}x\sim Q^{-4}\lambda^{-2-2b}=Q^{-4}\lambda^{-2-2b}$. There are not least two Glauber gluons and one soft gluon in these couplings. Infrared power counting for combination of these fields reads $\lambda^{3+b}=\lambda^{4}$. There is an additional gluon field or momentum operator in these couplings according to Lorentz invariance. Power clouting for the gluon field reads $\lambda^{1+\frac{b}{2}}=\lambda^{3/2}$ or $\lambda$. That of the momentum operator reads $\lambda$ or $\lambda^{2}$. Thus power counting for these couplings reads $\lambda$ or higher.

For couplings between Glauber gluon fields $A_{nG}$ and soft gluons without Glauber gluon fields $A_{\bar{n}G}$, the integral volume scales as $\int \ud^{4}x\sim Q^{-4}\lambda^{-4}$. In these couplings, there are at least two soft gluons and one Glauber gluon. Infrared power counting for combination of these fields reads $\lambda^{3+\frac{b}{2}}$. There is an additional gluon field or momentum operator in these couplings according to Lorentz invariance. Infrared power counting for the additional gluon field reads $\lambda^{1+\frac{b}{2}}$ or $\lambda$. That of the momentum operator reads $\lambda$ or $\lambda^{2}$. Thus infrared power counting for these couplings reads $\lambda^{\frac{b}{2}}$ or higher.

For couplings between Glauber gluons and soft fermions, the integral volume scales as $\int \ud^{4}x\sim Q^{-4}\lambda^{-4}$. In these couplings, there are at least two soft fermions and one Glauber gluon. Infrared power counting for combination of these fields reads $\lambda^{4+\frac{b}{2}}$. Momenta of these fields are quite small and there are not energy scale which may produce minus power of $\lambda$. Thus infrared power counting for these couplings reads $\lambda^{\frac{b}{2}}$ or higher.

For couplings between Glauber gluon fields $A_{nG}$ and fermions collinear to $n^{\mu}$, the integral volume scales as $\int \ud^{4}x\sim Q^{-4}\lambda^{-4}$. In these couplings, there are at least two collinear fermions and one Glauber gluon. Infrared power counting for combination of these fields reads $\lambda^{3+\frac{b}{2}}$.  Thus infrared power counting for these couplings reads $\lambda^{\frac{b}{2}-1}$ or higher.

For couplings between Glauber gluon fields $A_{nG}$ and gluons collinear to $n^{\mu}$, the integral volume scales as $\int \ud^{4}x\sim Q^{-4}\lambda^{-4}$. In these couplings, there are at least two collinear gluons and one Glauber gluon. Infrared power counting for combination of these fields reads $\lambda^{3+\frac{b}{2}}$. There is an additional gluon field or momentum operator in these couplings according to Lorentz invariance. Infrared power counting for the additional gluon field reads   $\lambda^{1+\frac{b}{2}}$ or $\lambda$. That of the momentum operator reads $\lambda^{0}$,  $\lambda$ or $\lambda^{2}$. Thus infrared power counting for these couplings reads $\lambda^{\frac{b}{2}-1}$ or higher.

For couplings between Glauber gluon fields $A_{nG}$ and fermions collinear to $\bar{n}^{\mu}$, the integral volume scales as $\int \ud^{4}x\sim Q^{-4}\lambda^{-2-b}$. In these couplings, there are at least one collinear fermion, one Glauber gluon and one fermion with momentum scales as $(n\cdot k, \bar{n}\cdot k,k_{n\perp})\sim Q(1,\lambda^{b},\lambda)$. Infrared power counting for combination of these fields reads $\lambda^{3+\frac{b}{2}}$.  Thus infrared power counting for these couplings reads $\lambda^{1-\frac{b}{2}}$ or higher.

For couplings between Glauber gluon fields $A_{nG}$ and gluons collinear to $\bar{n}^{\mu}$, the integral volume scales as $\int \ud^{4}x\sim Q^{-4}\lambda^{-2-b}$. In these couplings, there are at least one collinear gluon, one Glauber gluon and one gluon with momentum scales as $(n\cdot k, \bar{n}\cdot k,k_{n\perp})\sim Q(1,\lambda^{b},\lambda)$. Infrared power counting for combination of these fields reads $\lambda^{3+\frac{b}{2}}$ or higher. There is an additional gluon field or momentum operator in these couplings according to Lorentz invariance. Infrared power counting for these objects reads $\lambda^{0}$  or higher. Thus infrared power counting for these couplings reads $\lambda^{1-\frac{b}{2}}$ or higher.

Our results in this subsection are presented in  Table \ref{Power coupling}.
\begin{table}
\begin{tabular}{ccc}
\hline
Couplings & fields &power counting\\
\hline
Glauber gluons and ultrsoft gluons & ($A_{nG}$,$A_{us}$)& $\lambda$ or higher \\
\hline
Glauber gluons and ultrsoft fermions& ($A_{nG}$,$A_{us}$)& $\lambda^{\frac{3}{2}}$ or higher \\
\hline
Glauber gluons& ($A_{nG}$)& $\lambda^{\frac{b}{2}}$ or higher \\
\hline
Glauber gluons and soft gluons& ($A_{nG}$,$A_{\bar{n}G}$,$A_{s}$)& $\lambda$ or higher \\
\hline
Glauber gluons and soft gluons& ($A_{nG}$,$A_{s}$)& $\lambda^{\frac{b}{2}}$ or higher \\
\hline
Glauber gluons and soft fermions& ($A_{nG}$,$\psi_{s}$)& $\lambda^{\frac{b}{2}}$ or higher \\
\hline
Glauber gluons and collinear fermions& ($A_{nG}$,$\xi_{n}$)& $\lambda^{\frac{b}{2}-1}$ or higher \\
\hline
Glauber gluons and collinear gluons& ($A_{nG}$,$A_{n}$)& $\lambda^{\frac{b}{2}-1}$ or higher \\
\hline
Glauber gluons and collinear fermions& ($A_{nG}$,$\xi_{\bar{n}}$)& $\lambda^{1-\frac{b}{2}}$ or higher \\
\hline
Glauber gluons and collinear gluons& ($A_{nG}$,$A_{\bar{n}}$)& $\lambda^{1-\frac{b}{2}}$ or higher \\
\hline
\end{tabular}
\caption{Infrared power counting for couplings involving Glauber gluons, where $\bar{n}^{\mu}=\frac{1}{\sqrt{2}}(1,-\vec{n})$.}
\label{Power coupling}
\end{table}

\subsection{Leading power Lagrangian including Glauber gluons}
\label{LO-Lagrangian}

We consider leading power SCET Lagrangian including Glauber gluon fields in this subsection. There are two kings of SCET Lagrangian in literature, $\text{SCET}_{\text{\Rum{1}}}$ and $\text{SCET}_{\text{\Rum{2}}}$, which are suitable for studies on different observables. We do not distinguish here. For simplicity, we neglect couplings involving ultrasoft particles and couplings between Glauber gluons without soft gluons at first.

We start from a Glauber gluon $A_{nG}$ which couple to particles collinear to $n^{\mu}$.   Power counting for such coupling reads $\lambda^{\frac{b}{2}-1}$ according to results in Table \ref{Power coupling}.   The other end of the Glauber gluon may couple to ultrasoft particles, Glauber gluons,  soft particles or particles collinear to other directions. If the Glauber gluon couple to particles collinear to other directions at that end, then power counting for that coupling reads $\lambda^{1-\frac{b}{2}}$. The final power counting for couplings at two ends of the Glauber gluon reads $\lambda^{0}$.\footnote{There may be additional powers of $\lambda$ in practical diagrams even if  these diagrams  involve only  leading order couplings as shown in \cite{Rothstein:2016bsq}. This does not disturbs us here as we concern only power counting for  effective couplings in this paper.}

If the other end of the Glauber gluon involves soft particles,  then power counting for that coupling reads $\lambda^{\frac{b}{2}}$  or  $\lambda$.
For the former case, the final power counting for couplings at two ends of the Glauber gluon reads $\lambda^{b-1}\ge\lambda^{0}$. For the latter case, in which $b=1$, the other end of $A_{nG}$ involves a Glauber gluon of the type $A_{\bar{n}G}$. If the other end of $A_{\bar{n}G}$ couple to particles collinear to $\bar{n}^{\mu}$, then final power counting for these couplings reads $\lambda^{-\frac{1}{2}+1-\frac{1}{2}}=\lambda^{0}$. If the other end of $A_{\bar{n}G}$ couple to soft particles, then we can repeat the procedure and get the same result. In conclusion, the final power counting for these couplings reads $\lambda^{0}$ if there are not couplings between Glauber gluons and soft particles.

If a Glauber gluon is exchanged between soft particles, the power counting  for combination of couplings at the two ends of the diagram reads, $\lambda^{b}\ge \lambda^{1}$. We notice that minus power of $\lambda$ can only be produced by couplings between glauber gluons and collinear particles. Thus combination of  couplings involve Glauber gluons  power suppressed except for that couplings between glauber gluons and collinear particles are involved.

We then consider couplings involving ultrasoft particles and couplings between Glauber gluons without soft gluons. Power counting for couplings between Glauber gluons and ultrasoft particles reads $\lambda^{1}$ or $\lambda^{3/2}$.
Power counting for couplings between Glauber gluons without soft gluons reads $\lambda^{\frac{b}{2}}$.\footnote{One should not be confused with the possible fractional power in this coupling. There are other powers of $\lambda$ in  diagrams involving this coupling. For example, one consider the coupling between three Glauber gluons of the type $A_{nG}$. Two of them connect to a fermion collinear to $n^{\mu}$ and one of them collinear to $\bar{n}^{\mu}$. At leading power, two of them are of the type $\bar{n}\cdot A_{G}$ and one of them is of the type $n\cdot A_{G}$ in the coupling between these three Glauber gluons. According to Lorentz covariance, there is momentum term of the type $n\cdot k\sim \lambda^{2}$ instead of the type $k_{\perp}\sim \lambda$, which is produced by the coupling between the three Glauber gluons.  Thus power counting for combination of these couplings reads  $\lambda^{\frac{b}{2}-1}\lambda^{\frac{b}{2}-1}\lambda^{1-\frac{b}{2}}\lambda^{\frac{b}{2}}\lambda^{2-1}=\lambda^{b}$.}
According to analyses in above paragraphs, combination of other couplings involving Glauber gluons does not produce minus powers of $\lambda$.  Thus couplings involving ultrasoft particles and couplings between Glauber gluons without soft gluons are power suppressed.

According to above discussions, we see that:(1)Glauber gluons are exchanged between collinear particles and soft particles or between collinear particles at leading power;(2)there may be intermediate couplings between Glauber gluons and soft particles in these exchanges at leading power;(3)couplings between Glauber gluons and ultrasoft particles are power suppressed;(4)couplings between Glauber gluons without soft gluons are power suppressed. This is compatible with results in \cite{Rothstein:2016bsq}.

The leading power effective action can then be written as,
\begin{eqnarray}
\label{effective action}
I_{eff}&=&\sum_{n}I_{n}^{G}+I_{s}^{G}+I_{us}\\
I_{n}^{G}&=&\int\ud^{4} x\bar{\xi}_{n,p^{\prime}}\{in\cdot D+gn\cdot (A_{n,q}+A_{G,q})
\nonumber\\
&&+(\not\!\mathcal{P}_{\perp}+g\not\!A_{n,q}^{\perp})W_{n}\frac{1}{\bar{\mathcal{P}}}W_{n}^{\dag}
(\not\!\mathcal{P}_{\perp}+g\not\!A_{n,q^{\prime}}^{\perp})\}\frac{\bar{\not\!n}}{2}\xi_{n,p}\nonumber\\
&&+\int\ud^{4}x\frac{1}{2g^{2}}tr\{[i\mathcal{D}_{G}^{\mu}+g A_{n,q}^{\mu},i\mathcal{D}_{G}^{\nu}+g A_{n,q^{\prime}}^{\nu}]\}^{2}\nonumber\\
&&+\int\ud^{4}x2tr\{\bar{c}_{n,p^{\prime}}[i\mathcal{D}_{G\mu},[i\mathcal{D}_{G}^{\mu}+g A_{n,q}^{\mu},c_{n,p}]]\}\nonumber\\
&&+\int\ud^{4}x tr\{[i\mathcal{D}_{G\mu},A_{n,q}^{\mu}]
[i\mathcal{D}_{G\nu},A_{n,q^{\prime}}^{\nu}]\}
\\
I_{s}^{G}&=&\int\ud^{4}x\bar{q}_{s,p^{\prime}}(\not\!\mathcal{P}+g\not\!A_{s,q}+g\not\!A_{G,q})q_{s,p}
-\frac{1}{2}\int\ud^{4}x tr\{G_{s}^{\mu\nu}G_{\mu\nu}^{s}\}
\\
I_{us}&=&\int\ud^{4}x\bar{\psi}_{us}\not\!D \psi_{us}-\int\ud^{4}x\frac{1}{2}\{G_{\mu\nu}G^{\mu\nu}\},
\end{eqnarray}
where $\xi_{n,p}$, $A_{n,p}$, $c_{n,p}$(ghost) are collinear fields with momenta $p$, $\psi_{s,p}$ and $A_{s,p}$ are soft fields with momenta $p$, $\psi_{us}$ and $A_{us}$ are ultrasoft fields and
\begin{eqnarray}
W_{n}(x)&=&P\exp(ig\int_{-\infty}^{0}\ud s\bar{n}\cdot A_{n}(x+s\bar{n}))\\
D^{\mu}&=&\partial^{\mu}-i g A_{us}^{\mu}\\
\mathcal{P}^{\mu}(\phi_{q_{1}}^{\dag}\cdot\cdot\cdot\phi_{q_{m}}^{\dag}\phi_{p_{1}}\cdot\cdot\cdot\phi_{p_{n}}) &=&(p_{1}^{\mu}+\cdots+p_{n}^{\mu}-q_{1}^{\mu}\cdots-q_{m}^{\mu})
\nonumber\\
&&(\phi_{q_{1}}^{\dag}\cdot\cdot\cdot\phi_{q_{m}}^{\dag}\phi_{p_{1}}\cdot\cdot\cdot\phi_{p_{n}})\\
i\mathcal{D}^{\mu}&=&\frac{n^{\mu}}{2}\bar{n}\cdot \mathcal{P}+\mathcal{P}_{\perp}^{\mu}+\frac{\bar{n}^{\mu}}{2}i n\cdot D\\
i G_{s}^{\mu\nu}&=&[\mathcal{P}^{\mu}+g A_{s,q}^{\mu}+g A_{G,k}^{\mu},\mathcal{P}^{\nu}+g A_{s,q^{\prime}}^{\nu}+g A_{G,k^{\prime}}^{\nu}]\\
G^{\mu\nu}&=&\frac{i}{g}[D^{\mu},D^{\nu}]
\nonumber\\
\mathcal{D}_{G}^{\mu}&=&\mathcal{D}^{\mu}-\frac{\bar{n}^{\mu}}{2}ign\cdot (A_{G,q}),
\end{eqnarray}
summation over the collinear and soft momenta is understood implicitly. Some power suppressed terms tare added into the actio to maintain gauge covariance.
We see that diagrams involving $A_{nG}(k)$ rely on $n\cdot k$ through various propagators and vertexes involving $A_{nG}(k)$ are independent of $n\cdot k$ at leading power. This is crucial in our following discussions.

\section{Elastic Scattering Effects in Hadron Collisions}
\label{sspectator}

In this section, we consider kinematic effects of Glauber gluons in process inclusive enough.  These discussions are independent of details of couplings between Glauber gluons and collinear particles. We prove the cancellation of effects of final state interactions in inclusive process in this section.

Processes considered here can be written as,
\begin{equation}
H_{1}(P)+H_{2}(\bar{P})\to l^{+}l^{-}(q)+X
\end{equation}
or
\begin{equation}
H_{1}(P)+H_{2}(\bar{P})\to H_{3}(P_{3})+H_{4}(P_{4})+X
\end{equation}
with $P_{3}+P_{4}=q$, where $H_{1}$ and $H_{2}$ represent with momenta $P$ and $\bar{P}$, $l^{+}l^{-}(q)$ represents lepton pair with momentum $q$,   $H_{3}$ and $H_{4}$ represent detected final hadrons with momenta $P_{3}$ and $P_{4}$, $X$ represents any other states. We work in the center of mass frame of initial hadrons. In processes considered here, $q$ is a hard, $q^{2}\gg \Lambda_{QCD}^{2}$. At leading power of $\Lambda_{QCD}/(q^{2})^{1/2}$, $P$ and $\bar{P}$ are light like. Without loss of generality, we assume that $P^{\mu}$ is $+$-collinear and $\bar{P}^{\mu}$ is $-$-collinear.

In these processes, the hard subprocess is caused by a hard vertex, which is denoted as $J(x)$. For example, the hard electromagnetic vertex in $\text{SCET}_{\text{\Rum{2}}}$   takes the form\cite{Rothstein:2016bsq},
\begin{equation}
\label{emhaver}
J=\bar{\xi}_{n}W_{n}S_{n}^{\dag}\Gamma S_{\bar{n}}W_{\bar{n}}^{\dag}\xi_{\bar{n}},
\end{equation}
where $W_{n}$ and $W_{\bar{n}}$ are Wilson lines of collinear gluons, $S_{n}$ and $S_{\bar{n}}$ are Wilson lines of soft gluons.

We do not consider quantities dependent on $q_{\perp}$ in this paper. Thus one can integrate out $q_{\perp}$ in following discussion. For simplicity, we work in the Feynman gauge here.

\subsection{Elastic scattering effects without interactions between spectators and active particles}
\label{ss0as}

\begin{figure*}
\begin{center}
\includegraphics[scale=0.5]{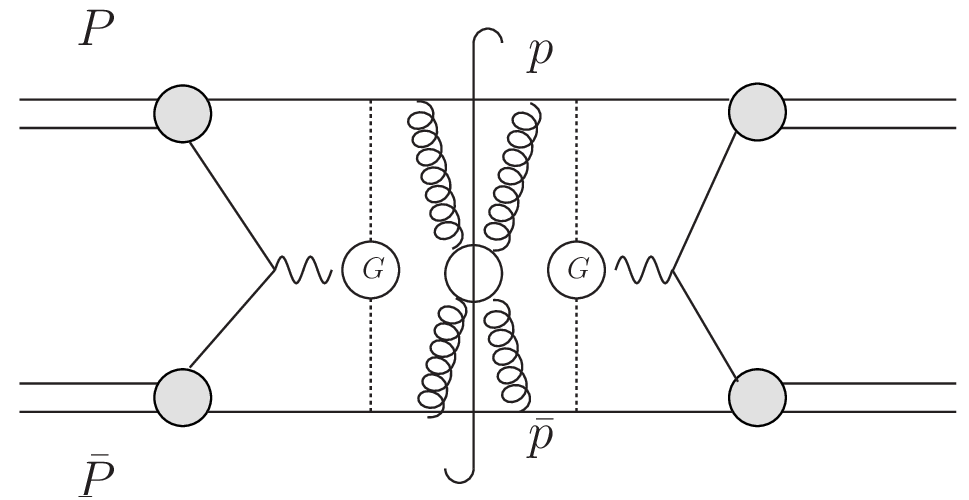}
\end{center}
\caption{Example of diagrams without coherence between spectators and active particles except for the vertex which creta spectators and active particles from initial particles.}
\label{ss2}
\end{figure*}

In this subsection, we start from spectator-spectator interactions and  neglect interactions between spectators and active particles.  According to results in \cite{Rothstein:2016bsq}, summations over ladder diagrams shown in Fig.\ref{ss} cancel out in processes inclusive enough. Such cancellation originates from unitarity and it is independent of details of couplings between Glauber gluons and collinear particles according to discussions in this subsection. In addition, we see that such cancellation also works for non-Glauber interactions  like that  shown in \ref{ss2}  given that spectator-active type interactions are neglected.

Let us start from diagrams shown in Fig.\ref{ss}. The Diagrams shown in \ref{ss} can be written as,
\begin{eqnarray}
\label{intss}
S(P,\bar{P},p,\bar{p})&\equiv& \int\ud^{4} x_{1}\int\ud^{4} x_{2}\int\ud^{4} x_{3}\int\ud^{4} x_{4}\int\ud^{4}z e^{i(P+\bar{P}-p-\bar{p})\cdot z}
\nonumber\\
&&\big<P\bar{P}|\bar{T}\{\mathcal{O}^{\dag}(x_{1})\bar{\mathcal{O}}^{\dag}(x_{2})J^{\dag}(z)U^{\dag}(\infty,-\infty)\}|p\bar{p}\big>
\nonumber\\
&&\big<p\bar{p}|T\{\mathcal{O}(x_{3})\bar{\mathcal{O}}(x_{4})J(0)U(\infty,-\infty)\}|P\bar{P}\big>,
\end{eqnarray}
where $T$and $\bar{T}$ represent time order and anti-time order operators, $J(x)$ represents the hard vertexes that annihilate active particles, $O(x)$($\bar{O}(x)$) represents the vertex that produce active particles and spectators from the initial particle $P$($\bar{P}$), $U(t_{1},t_{2})$ represents the time evolution operator of the effective theory in the interaction picture.

If we neglect interaction terms between active particle fields and spectator fields in the effective theory, then the effective action can be written as
\begin{equation}
I_{eff}=I_{ac}(\psi_{ac},A_{ac}^{\mu})+I_{sp}(\psi_{sp},A_{sp}^{\mu}),
\end{equation}
where $I_{ac}$ is the part of $I_{eff}$ that describes  active particles and $I_{sp}$ is the part of $I_{eff}$ that describes spectators. Interaction terms between active particle fields and spectator fields have been dropped in above decomposition. In this case, we have
\begin{equation}
\label{timefactorization}
U(t_{1},t_{2})=U_{ac}(t_{1},t_{2})U_{sp}(t_{1},t_{2})=U_{sp}(t_{1},t_{2})U_{ac}(t_{1},t_{2}),
\end{equation}
where $U_{ac}(t_{1},t_{2})$ and $U_{sp}(t_{1},t_{2})$ represent the time evolution operator corresponding to $I_{ac}$ and $I_{sp}$ respectively.

We consider the Wick contractions of fields in (\ref{intss}). Fields in $U_{sp}$ does not contrate with those in the current $J$ as $J$ is functional of active particle fields. We notice that energies of spectators in Fig.\ref{ss} flow out of the vertexes $\mathcal{O}$ and $\bar{\mathcal{O}}$.  Hence couplings involving spectators all occur after the production of spectators at the effective vertexes. Especially, interactions induced by vertexes in $U_{sp}$ should occur after those induced by the vertexes ${\mathcal O}$($\bar{\mathcal{O}}$) once one neglect couplings between active particles and (ultra)soft or Glauber gluons in $U_{sp}$. That is to say, contractions between fields in ${\mathcal O}$($\bar{\mathcal{O}}$) and those in $U_{sp}$ cancel out  unless time coordinates of fields in ${\mathcal O}$($\bar{\mathcal{O}}$) are smaller than those in $U_{sp}$. We have,
\begin{eqnarray}
&&\big<P\bar{P}|\bar{T}\{\mathcal{O}^{\dag}(x_{1})\bar{\mathcal{O}}^{\dag}(x_{2})J^{\dag}(z)U^{\dag}(\infty,-\infty)\}|p\bar{p}\big>
\nonumber\\
&=&\big<P\bar{P}|\bar{T}\{\mathcal{O}^{\dag}(x_{1})\bar{\mathcal{O}}^{\dag}(x_{2})J^{\dag}(z)U_{ac}^{\dag}(\infty,-\infty)\}
U_{sp}^{\dag}(\infty,-\infty)|p\bar{p}\big>
\\
&&\big<p\bar{p}|T\{\mathcal{O}(x_{3})\bar{\mathcal{O}}(x_{4})J(0)U(\infty,-\infty)\}|P\bar{P}\big>
\nonumber\\
&=&\big<p\bar{p}|U_{sp}(\infty,-\infty)T\{\mathcal{O}(x_{3})\bar{\mathcal{O}}(x_{4})J(0)U_{ac}(\infty,-\infty)\}|P\bar{P}\big>.
\end{eqnarray}

We then consider evolution of the final state $|p\bar{p}\big>$ under $U_{sp}(\infty,-\infty)$. Elastic scattering processes between spectators exchanges transverse momenta, colors and angular momenta between these particles. Total momentum, color and angular momentum  of these particles do not change in processes. Thus all possible pairs $|p\bar{p}\big>$ with fixed total  momenta, color and spin form the invariant subspace of $U_{sp}$ given that one neglects inelastic scattering processes between spectators as in Fig.\ref{ss}. That is,
\begin{eqnarray}
&&\sum\int\ud^{2}\Delta p_{\perp}
U_{sp}^{\dag}(\infty,-\infty)|p\bar{p}\big>\big<p\bar{p}|U_{sp}(\infty,-\infty)
\nonumber\\
&=&\sum\int\ud^{2}\Delta p_{\perp}
|p\bar{p}\big>\big<p\bar{p}|U_{sp}^{\dag}(\infty,-\infty)U_{sp}(\infty,-\infty)
\nonumber\\
&=&\sum\int\ud^{2}\Delta p_{\perp}
|p\bar{p}\big>\big<p\bar{p}|,
\end{eqnarray}
where the summation is made over all possible color and angular momentum distributions of the pair $p\bar{p}$ with fixed total color and angular momentum, $\Delta p_{\perp}$ is defined as,
\begin{equation}
\Delta p_{\perp}\equiv p_{\perp}-\bar{p}_{\perp}.
\end{equation}
We then have,
\begin{eqnarray}
&&\sum\int\ud^{2}\Delta p_{\perp} S(P,\bar{P},p,\bar{p})
\nonumber\\
&\equiv&\sum\int\ud^{2}\Delta p_{\perp} \int\ud^{4} x_{1}\int\ud^{4} x_{2}\int\ud^{4} x_{3}\int\ud^{4} x_{4}\int\ud^{4}z e^{i(P+\bar{P}-p-\bar{p})\cdot z}
\nonumber\\
&&\big<P\bar{P}|\bar{T}\{\mathcal{O}^{\dag}(x_{1})\bar{\mathcal{O}}^{\dag}(x_{2})J^{\dag}(z)U_{ac}^{\dag}(\infty,-\infty)\}|p\bar{p}\big>
\nonumber\\
&&\big<p\bar{p}|T\{\mathcal{O}(x_{3})\bar{\mathcal{O}}(x_{4})J(0)U_{ac}(\infty,-\infty)\}|P\bar{P}\big>.
\end{eqnarray}
Thus elastic scattering effects between spectators cancel out in processes inclusive enough given that interactions terms between spectators and other particles haven been dropped out of the effective theory as in Fig.\ref{ss}. Such cancellation is independent of details of elastic scattering interactions between spectators.

The cancelation can be generalized to effective theories with inelastic interactions involving spectators but without coherence between spectators and active particles. In this case, spectators may emission real final particles. Thus summation over all possible final states is necessary. That is,
\begin{eqnarray}
\sum_{p,\bar{p},\ldots,}
U_{sp}^{\dag}(\infty,-\infty)|p\bar{p}\ldots\big>\big<p\bar{p}\ldots |U_{sp}(\infty,-\infty)
&=&\sum_{p,\bar{p},\ldots,}
|p\bar{p}\ldots \big>\big<p\bar{p}\ldots|,
\end{eqnarray}
where the summation is made over all possible final particles, momentum, color and angular momentum distributions with fixed total momentum, color and angular momentum.
Thus effects of interactions involving spectators cancel out  in  processes inclusive enough given that possible coherence between spectators and active particles haven been dropped out of the effective theory as in Fig.\ref{ss2}.

\subsection{Cancellation of effects of interactions after the hard collision}
\label{ficanel}

In this subsection, we take into account active-spectator type interactions and prove the cancellation of of effects of final state interactions in  process inclusive enough. Such cancellation is the direct result of unitarity and independent of details of interactions between particles.

\begin{figure*}
\begin{center}
\includegraphics[scale=0.5]{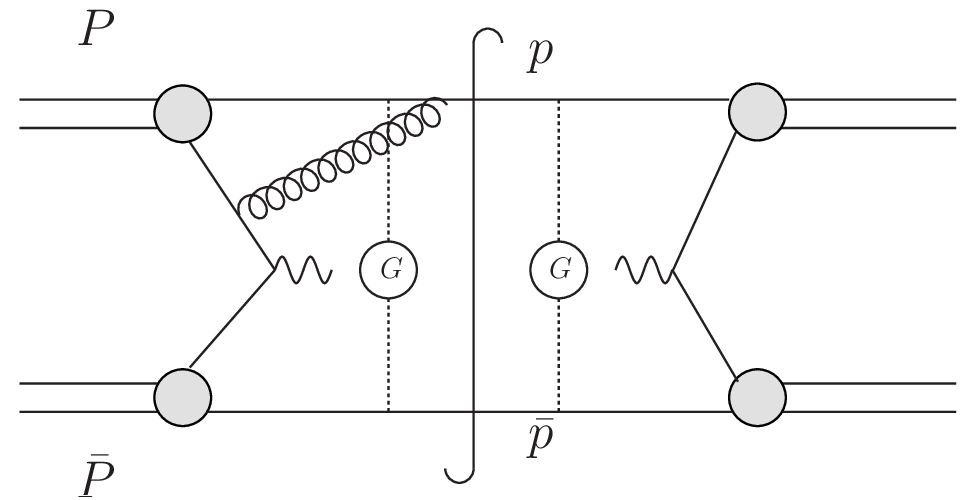}
\end{center}
\caption{Example of diagrams with interactions between spectators and active particles.}
\label{ssas}
\end{figure*}

For diagrams with interactions between spectators and active particles, like those shown in Fig.\ref{ssas} and their conjugations, the situation is more complicated.
Let's consider the following quantity,
\begin{eqnarray}
\label{intg}
\mathcal{H}(P,\bar{P},q^{+},q^{-})
&\equiv& \sum_{X}\int\frac{\ud^{2}q_{\perp}}{(2\pi)^{2}}\int\ud^{4} x_{1}\int\ud^{4} x_{2}\int\ud^{4} x_{3}\int\ud^{4} x_{4}\int\ud^{4}z e^{iq\cdot z}
\nonumber\\
&&\big<P\bar{P}|\bar{T}\{\mathcal{O}^{\dag}(x_{1})\bar{\mathcal{O}}^{\dag}(x_{2})J^{\dag}(z)U^{\dag}(\infty,-\infty)\}|H(q)X\big>
\nonumber\\
&&\big<H(q)X|T\{\mathcal{O}(x_{3})\bar{\mathcal{O}}(x_{4})J(0)U(\infty,-\infty)\}|P\bar{P}\big>
,
\end{eqnarray}
where $H(q)$ represents the detected lepton pair or hadron pair with total momenta $q$, the summation is made over all possible final states. We see that $\mathcal{H}$ describes the hadronic part of the cross section. Generally speaking, there are interaction terms between active particles in the time evolution operator $U(t_{1},t_{2})$ and the factorization (\ref{timefactorization}) does not work in this case.

As in (\ref{intss}), contraction between a field in $\mathcal{O}$($\bar{\mathcal{O}}$) and that in other operators occurs only if the time coordinate of the field in $O(x)$ smaller than that of the field in other operators. Other wise the contraction does not contributes to $\mathcal{H}$. Thus we can write $\mathcal{H}$ as
\begin{eqnarray}
\mathcal{H}(P,\bar{P},q^{+},q^{-})
&\equiv& \sum_{X}\int\frac{\ud^{2}q_{\perp}}{(2\pi)^{2}}\int\ud^{4} x_{1}\int\ud^{4} x_{2}\int\ud^{4} x_{3}\int\ud^{4} x_{4}\int\ud^{4}z e^{iq\cdot z}
\nonumber\\
&&\big<P\bar{P}|\mathcal{O}^{\dag}(x_{1})\bar{\mathcal{O}}^{\dag}(x_{2})\bar{T}\{J^{\dag}(z)U^{\dag}(\infty,-\infty)\}|H(q)X\big>
\nonumber\\
&&\big<H(q)X|T\{J(0)U(\infty,-\infty)\}\mathcal{O}(x_{3})\bar{\mathcal{O}}(x_{4})|P\bar{P}\big>,
\end{eqnarray}
 where the order between operators $\mathcal{O}$ and $\bar{\mathcal{O}}$ does not affect the result as contractions between fields in them vanish.

We notice that,
\begin{eqnarray}
T\{J(x)U(\infty,-\infty)\}&=&U(\infty,x^{0})J(x)U(x^{0},-\infty)
\nonumber\\
&=&U(\infty,\max\{x^{0},0\})U(\max\{x^{0},0\},x^{0})J(x)U(x^{0},-\infty)
\end{eqnarray}
and have,
\begin{eqnarray}
&&\mathcal{H}(P,\bar{P},q^{+},q^{-})
\nonumber\\
&\equiv& \sum_{X}\int\frac{\ud^{2}q_{\perp}}{(2\pi)^{2}}\int\ud^{4} x_{1}\int\ud^{4} x_{2}\int\ud^{4} x_{3}\int\ud^{4} x_{4}\int\ud^{4}z e^{iq\cdot z}
\nonumber\\
&&\big<P\bar{P}|\mathcal{O}^{\dag}(x_{1})\bar{\mathcal{O}}^{\dag}(x_{2})\bar{T}\{J^{\dag}(z)U^{\dag}(\max\{x^{0},0\},-\infty)\}
U^{\dag}(\infty,-\max\{x^{0},0\})|H(q)X\big>
\nonumber\\
&&\big<H(q)X|U(\infty,\max\{x^{0},0\})T\{J(0)U(\max\{x^{0},0\},-\infty)\}\mathcal{O}(x_{3})\bar{\mathcal{O}}(x_{4})|P\bar{P}\big>.
\end{eqnarray}
The summation in above quantity is made over all possible final states. As a result, completeness of the final states $X$  and unitarity of the time evolution operator $U(t_{1},t_{2})$ hint that
\begin{eqnarray}
&&\sum_{X}\int\frac{\ud^{2}q_{\perp}}{(2\pi)^{2}}U^{\dag}(\infty,-\max\{x^{0},0\})|H(q)X\big>\big<H(q)X|U(\infty,\max\{x^{0},0\})
\nonumber\\
&=&\sum_{X}\int\frac{\ud^{2}q_{\perp}}{(2\pi)^{2}}|H(q)X\big>\big<H(q)X|U^{\dag}(\infty,-\max\{x^{0},0\})U(\infty,\max\{x^{0},0\})
\nonumber\\
&=&\sum_{X}\int\frac{\ud^{2}q_{\perp}}{(2\pi)^{2}}|H(q)X\big>\big<H(q)X|.
\end{eqnarray}
We have,
\begin{eqnarray}
\label{fincanH}
&&\mathcal{H}(P,\bar{P},q^{+},q^{-})
\nonumber\\
&\equiv& \sum_{X}\int\frac{\ud^{2}q_{\perp}}{(2\pi)^{2}}\int\ud^{4} x_{1}\int\ud^{4} x_{2}\int\ud^{4} x_{3}\int\ud^{4} x_{4}\int\ud^{4}z e^{iq\cdot z}
\nonumber\\
&&\big<P\bar{P}|\mathcal{O}^{\dag}(x_{1})\bar{\mathcal{O}}^{\dag}(x_{2})\bar{T}\{J^{\dag}(z)U^{\dag}(\max\{x^{0},0\},-\infty)\}
|H(q)X\big>
\nonumber\\
&&\big<H(q)X|T\{J(0)U(\max\{x^{0},0\},-\infty)\}\mathcal{O}(x_{3})\bar{\mathcal{O}}(x_{4})|P\bar{P}\big>.
\end{eqnarray}
Hence interactions after the hard collision cancel out in the $\mathcal{H}(P,\bar{P},q^{+},q^{-})$.  This is compatible with results in \cite{B:1985,CSS:1985,CSS:1988}.

Especially, effects of couplings between Glauber gluons and particles produced by the hard vertex cancel out in $\mathcal{H}$ as these couplings occur after the time $0$ or $z^{0}$ and $z^{0}\sim 0\sim 1/(q^{2})^{1/2}$.

\section{Glauber gluons coupling to active particles}
\label{AG}

In this section, we consider Glauber gluons coupling to active particles. These Glauber gluons are absorbed into collinear or soft Wilson lines in ladder diagrams of Glauber gluons in \cite{Rothstein:2016bsq}. We extent the conclusion to general situation in this section. Explicitly we see that, (1)effects of Glauber gluons exchanged between active and soft particles can be absorbed into soft Wilson lines in $\mathcal{H}(P,\bar{P},q^{+},q^{-})$;(2)effects of Glauber gluons exchanged between active particles can be absorbed into soft Wilson lines in $\mathcal{H}(P,\bar{P},q^{+},q^{-})$;(3)effects of Glauber gluons exchanged between active particles and spectators can be absorbed into collinear Wilson lines in $\mathcal{H}(P,\bar{P},q^{+},q^{-})$.

Glauber gluons look like special collinear or soft gluons. For example, Glauber gluons $A_{nG}^{\mu}$ can be viewed as collinear gluons $A_{n,p}^{\mu}$ with $\bar{n}\cdot p=0$ or soft gluons $A_{s,q}^{\mu}$ with $n\cdot q=0$.  As we always integral over all loop momenta region in practical diagrams, distinguishing different modes in loop integrals is quite technical.   In the frame of perturbative QCD, momenta of gluons coupling to active particles are not pinched in Glauber region\cite{B:1985,CSS:1985,CSS:1988}. As a result, one can deform the integral contour to avoid the Glauber region of gluons coupling to active particles. In effective theory like SCET, one needs systematic subtraction schemes of which the details are beyond the scope of this paper to distinguish different modes. We emphasize that whatever the subtraction scheme is, the eikonal approximation is important while dealing with couplings between collinear and soft(ultrasoft) particles.

The key point is the eikonalization of couplings between Glauber gluons and active particles at leading power of $\lambda$ and $\eta$. As a sequence, Glauber gluons behave like soft or collinear gluons while coupling to active particles. This provides us with the possibility to absorb Glauber gluons coupling to active particles into definition of soft or collinear gluons. Even if one does not concern the definition of different modes, absorption of these Glauber gluons into soft or collinear Wilson lines is the direct result of eikonlization as presented in following subsections.

We do not consider couplings between Glauber gluons and final active particles as final interactions cancel out in $\mathcal{H}(P,\bar{P},q^{+},q^{-})$(\ref{fincanH}).  Without loss of generality, we consider effects of Glauber gluons coupling to $+$-collinear active particles in this section.  Specifically, we consider couplings between some glauber gluons $l_{1},\ldots, l_{n}$ and a $+$-collinear active particle $k$ in following subsections.

\subsection{Eikonal approximation in couplings between collinear particles and soft (ultrasoft) gluons}
\label{Eik}

In this subsection, we briefly explain the eikonal approximation in couplings between collinear particles and soft(ultrasoft) gluons.\footnote{Couplings between collinear particles and soft(ultrasoft) fermions are power suppressed\cite{SCET:2001,SCET:2002}.} Let us consider coupling between a $+$-collinear particle $p$ and a soft(ultrasoft) gluon $q$.
Power counting for $p$ and $q$ reads
\begin{equation}
(p^{+}, p^{-}, p_{\perp})\sim Q(1,\lambda^{2},\lambda),\quad
q^{\mu}\sim Q\lambda(\text{for soft gluons })\quad \text{or}\quad  q^{\mu}\sim Q\lambda^{2}(\text{for ultrasoft gluons }).
\end{equation}
We have
\begin{equation}
(p\pm q)^{2}=p^{2}\pm 2p\cdot q+q^{2}\simeq p^{2}\pm 2p^{+} q^{-}.
\end{equation}
According to Lorentz invariance the gluon field $A^{\mu}(q)$ should contract with some vectors. In the limit $\lambda\to 0$, $p$ behaves like a 1-dimensional particle and there is only one direction relevant to $p$(the plus direction) in this limit. Hence $A^{\mu}(q)$ should contract with the plus direction in the limit $\lambda\to 0$. In other words, $A^{\mu}(q)$ should contract with vectors collinear to the plus direction at leading power of $\lambda$. This conforms with the leading power action(\ref{LO-Lagrangian}). In summary, we can make the approximation
\begin{equation}
\label{eikapp}
(p\pm q)^{2}\simeq p^{2}\pm 2p^{+} q^{-}\quad A^{\mu}\equiv (A^{+},A^{-},A_{\perp })\simeq (A^{+},0,0)
\end{equation}
in couplings between soft(ultrasoft) gluons and collinear particles.
The approximation (\ref{eikapp}) is termed as eikonal approximation in literature.

It is interesting to consider the coordinate space version of approximation (\ref{eikapp}),
\begin{eqnarray}
A_{s}^{\mu}(x)&\equiv&(A_{s}^{+}(x),A_{s}^{-}(x),A_{s\perp}(x)) \simeq  (A_{s}^{+}(x^{+},0,0),0,0)
\nonumber\\
A_{us}^{\mu}(x)&\equiv& (A_{us}^{+}(x),A_{us}^{-}(x),A_{us\perp}(x))\simeq  (A_{us}^{+}(x^{+},0,0),0,0).
\end{eqnarray}
We have\footnote{According to the formula $\theta(x-y)=\int\frac{\ud k}{2\pi }\frac{i e^{-ik(x-y)}}{k+i\epsilon}$, one has $(\partial_{x}+\epsilon)\theta(x-y)=\delta(x-y)$ and $(\partial_{x}-\epsilon)\theta(y-x)=-\delta(y-x)$.}
\begin{eqnarray}
&&\partial^{\mu}-ig (A_{s}^{+}(x^{+},0,0),0,0)-ig (A_{s}^{+}(x^{+},0,0),0,0)-\epsilon
\nonumber\\
&=&
\left(\mathcal{P}\exp\left(ig\int_{0}^{\infty}\ud s (A_{s}^{+}+A_{us}^{+})(x^{+}+s,0,0) \right)\right)^{\dag}
\nonumber\\
&&
(\partial^{\mu}-\epsilon)\mathcal{P}\exp\left(ig\int_{0}^{\infty}\ud s (A_{s}^{+}+A_{us}^{+})(x^{+}+s,0,0) \right)
\\
&&\partial^{\mu}-ig (A_{s}^{+}(x^{+},0,0),0,0)-ig (A_{s}^{+}(x^{+},0,0),0,0)+\epsilon
\nonumber\\
&=&\mathcal{P}\exp\left(ig\int_{-\infty}^{0}\ud s (A_{s}^{+}+A_{us}^{+})(x^{+}+s,0,0) \right)
\nonumber\\
&&
(\partial^{\mu}+\epsilon)\left(\mathcal{P}\exp\left(ig\int_{-\infty}^{0}\ud s (A_{s}^{+}+A_{us}^{+})(x^{+}+s,0,0) \right)\right)^{\dag}.
\end{eqnarray}
That is to say, absorption of soft and ultrasolt gluons into light like Wilson lines is the direct result of the eikonal approximation.\footnote{We do not distinguish Wilson lines of soft and ultrasoft gluons here as it is irrelevant to the main result in this subsection.}

The approximation (\ref{eikapp}) also work in couplings between $+$-collinear particles and $-$-collinear gluons. However, couplings between Glauber gluons and collinear particles are more subtle. Take the coupling between a $+$-collinear particle $p$ and a Glauber gluon $l$ as an example. Power counting for $p$ and $l$ reads
\begin{equation}
(p^{+}, p^{-}, p_{\perp})\sim Q(1,\lambda^{2},\lambda),\quad
(l^{+}, l^{-}, l_{\perp})\sim Q(\lambda^{b},\lambda^{2},\lambda)(b=1,2).
\end{equation}
We have
\begin{equation}
2p^{+}l^{-}\sim 2p\cdot l\sim l^{2}
\end{equation}
and the approximation
\begin{equation}
(p\pm l)^{2}\simeq p^{2}+2p^{+}l^{-}
\end{equation}
does not work here. In other words couplings between Glauber gluons and collinear particles does not simply eikonalize. This prevent the absorption of  Glauber gluons into Wilson lines independent of explicit diagrams.

\subsection{Eikonalization of couplings between active particles and Glauber gluons in $\mathcal{H}(P,\bar{P},q^{+},q^{-})$}
\label{CAG}

In this subsection, we consider couplings between Glauber gluons and active particles in $\mathcal{H}(P,\bar{P},q^{+},q^{-})$ and prove the eikonalization of these couplings.

\begin{figure*}
\begin{center}
\includegraphics[scale=0.5]{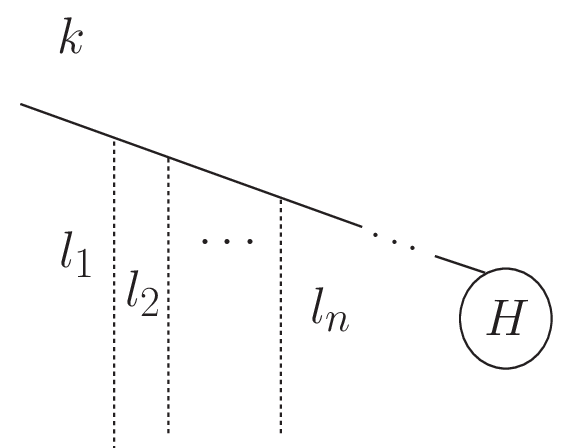}
\end{center}
\caption{Couplings between Glauber gluons and an active particle, where $H$ represents the hard vertex and dot lines represent Glauber gluons. Other parts of the whole diagram are not displayed explicitly.}
\label{CG+A}
\end{figure*}

Without loss of generality, we consider couplings between Glauber gluons and a $+$-collinear active particle $k$
\begin{equation}
|k^{+}|\gg |k_{\perp}|\gg |k^{-}|.
\end{equation}
Let us consider couplings between $k$ and some Glauber gluons $l_{1},\ldots,\l_{n}$ as shown in Fig.\ref{CG+A}.
At leading power of $\lambda$ and $\eta$, we have\footnote{If $k$ is a $+$-collinear spectator then momenta of $+$-collinear active particles depend on $(l_{1},\ldots, l_{n})$.
On the other hand, if $k$ is the active particle collinear to the plus direction  as considered here then momenta of $+$-collinear spectators are independent of $(l_{1},\ldots, l_{n})$.}
\begin{eqnarray}
Fig.\ref{CG+A}&=&\int\frac{\ud l_{1}^{-}}{2\pi}\cdots\int\frac{\ud l_{n}^{-}}{2\pi}
\int\frac{\ud^{D-2} l_{1\perp}}{(2\pi)^{D-2}}\cdots\int\frac{\ud^{D-2} l_{n\perp}}{(2\pi)^{D-2}}
\nonumber\\
&&\frac{1}{k^{-}+l_{1}^{-}+\frac{(k_{\perp}+l_{1\perp})^{2}}{2k^{+}}+i\epsilon}\cdots
\frac{1}{k^{-}+l_{1}^{-}+\cdots+l_{n}^{-}+\frac{(k_{\perp}+l_{1\perp}+\cdots+l_{n\perp})^{2}}{2k^{+}}+i\epsilon}
\nonumber\\
&&\cdots \frac{1}{k^{-}+l_{1}^{-}+\cdots+l_{n}^{-}+\cdots+\frac{(k_{\perp}+l_{1\perp}+\cdots+l_{n\perp}+\cdots)^{2}}{2k^{+}}+i\epsilon}
\nonumber\\
&&\frac{1}{l_{1\perp}^{2}+i\epsilon}\cdots
\frac{1}{l_{n\perp}^{2}+i\epsilon}
\times\mathcal{F} (l_{1\perp},\ldots, l_{n\perp})
\nonumber\\
&&\times \text{(terms independent of $l_{j}^{-}$ and $l_{j\perp}$)($1\le j\le n$)},
\end{eqnarray}
where $D=4-2\varepsilon$ and $\mathcal{F} (l_{1\perp},\ldots, l_{n\perp})$ represents terms depend on $(l_{1\perp},\ldots, l_{n\perp})$ on the other ends of these Glauber gluons.

One can easily verify that
\begin{eqnarray}
&&\int\frac{\ud l_{1}^{-}}{2\pi}\cdots\int\frac{\ud l_{n}^{-}}{2\pi}
\left(\frac{1}{k^{-}+l_{1}^{-}+\frac{(k_{\perp})^{2}}{2k^{+}}+i\epsilon}
\cdots \frac{1}{k^{-}+l_{1}^{-}+\cdots+l_{n}^{-}+\cdots+\frac{(k_{\perp}+\cdots)^{2}}{2k^{+}}+i\epsilon}\right.
\nonumber\\
&&\left. -\frac{1}{k^{-}+l_{1}^{-}+\frac{(k_{\perp}+l_{1\perp})^{2}}{2k^{+}}+i\epsilon}
\cdots \frac{1}{k^{-}+l_{1}^{-}+\cdots+l_{n}^{-}+\cdots+\frac{(k_{\perp}+l_{1\perp}+\cdots+l_{n\perp}+\cdots)^{2}}{2k^{+}}+i\epsilon}\right)
\nonumber\\
&=&0.
\end{eqnarray}
Hence
\begin{eqnarray}
\label{eikG+}
Fig.\ref{CG+A}&=&\int\frac{\ud l_{1}^{-}}{2\pi}\cdots\int\frac{\ud l_{n}^{-}}{2\pi}
\int\frac{\ud^{D-2} l_{1\perp}}{(2\pi)^{D-2}}\cdots\int\frac{\ud^{D-2} l_{n\perp}}{(2\pi)^{D-2}}
\nonumber\\
&&\frac{1}{k^{-}+l_{1}^{-}+\frac{(k_{\perp})^{2}}{2k^{+}}+i\epsilon}\cdots
\frac{1}{k^{-}+l_{1}^{-}+\cdots+l_{n}^{-}+\frac{(k_{\perp})^{2}}{2k^{+}}+i\epsilon}
\nonumber\\
&&\cdots \frac{1}{k^{-}+l_{1}^{-}+\cdots+l_{n}^{-}+\cdots+\frac{(k_{\perp}+\cdots)^{2}}{2k^{+}}+i\epsilon}
\nonumber\\
&&\frac{1}{l_{1\perp}^{2}+i\epsilon}\cdots
\frac{1}{l_{n\perp}^{2}+i\epsilon}
\times\mathcal{F} (l_{1\perp},\ldots, l_{n\perp})
\nonumber\\
&&\times \text{(terms independent of $l_{j}^{-}$ and $l_{j\perp}$)($1\le j\le n$)}
\end{eqnarray}
at leading power of $\lambda$ and $\eta$. That is to say, couplings between Glauber gluons and $k$ eikonalize at leading power of $\lambda$ and $\eta$.

For couplings between Glauber gluons and active particles collinear to other directions we have similar results. In conclusion, couplings between Glauber gluons and active particles eikonalize at leading power of $\lambda$ and $\eta$ in $\mathcal{H}(P,\bar{P},q^{+},q^{-})$.

\subsection{Glauber gluons exchanged between active and soft particles}
\label{ASG}

We consider effects of Glauber gluons exchanged between  active particles and soft particles in this subsection. These Glauber gluons can be absorbed into soft Wilson lines according to discussions here.

Without loss of generality, we consider Glauber gluons exchanged between $+$-collinear active particles and soft particles.  To specify our discussions, let us consider Glauber gluons($l_{1},\ldots,l_{n}$) exchanged between a $+$-collinear active particle $k$ and a soft particle $k_{s}$.

Couplings between $k$ and Glauber gluons read(\ref{eikG+})
\begin{eqnarray}
\label{ASG+}
&&
\int\frac{\ud^{D} l_{1}}{(2\pi)^{D}}\cdots\int\frac{\ud^{D} l_{n}}{(2\pi)^{D}}
\frac{1}{k^{-}+l_{1}^{-}+\frac{(k_{\perp})^{2}}{2k^{+}}+i\epsilon}\cdots
\frac{1}{k^{-}+l_{1}^{-}+\cdots+l_{n}^{-}+\frac{(k_{\perp})^{2}}{2k^{+}}+i\epsilon}
\nonumber\\
&&\cdots \frac{1}{k^{-}+l_{1}^{-}+\cdots+l_{n}^{-}+\cdots+\frac{(k_{\perp}+\cdots)^{2}}{2k^{+}}+i\epsilon}
\nonumber\\
&&\frac{1}{l_{1\perp}^{2}+i\epsilon}\cdots\frac{1}{l_{n\perp}^{2}+i\epsilon}
\times \text{(terms independent of $l_{j}$)($1\le j\le n$)}
\nonumber\\
&&\times\text{(terms on the other ends of $l_{1},\ldots,l_{n}$)}
\end{eqnarray}
at leading power of $\lambda$ and $\eta$, where $D=4-2\varepsilon$. We compare the result with the  case that $k$ couple to soft gluons $q_{1},\ldots,q_{n}$. After the eikonal approximation couplings between $q_{1},\ldots,q_{n}$ and $k$ read
\begin{eqnarray}
\label{ASS+}
&&\int\frac{\ud^{D} q_{1}}{(2\pi)^{D}}\cdots\int\frac{\ud^{D} q_{n}}{(2\pi)^{D}}
\frac{1}{k^{-}+q_{1}^{-}+\frac{(k_{\perp})^{2}}{2k^{+}}+i\epsilon}\cdots
\frac{1}{k^{-}+q_{1}^{-}+\cdots+q_{n}^{-}+\frac{(k_{\perp})^{2}}{2k^{+}}+i\epsilon}
\nonumber\\
&&\cdots \frac{q}{k^{-}+q_{1}^{-}+\cdots+q_{n}^{-}+\cdots+\frac{(k_{\perp}+\cdots)^{2}}{2k^{+}}+i\epsilon}
\nonumber\\
&&\frac{1}{q_{1}^{+}q_{1}^{-}+q_{1\perp}^{2}+i\epsilon}\cdots\frac{1}{q_{n}^{+}q_{n}^{-}+q_{n\perp}^{2}+i\epsilon}
\times \text{(terms independent of $q_{j}$)($1\le j\le n$)}
\nonumber\\
&&\times\text{(terms on the other ends of $q_{1},\ldots,q_{n}$)}
\end{eqnarray}
at leading power of $\lambda$ and $\eta$.  Comparing (\ref{ASG+}) with  (\ref{ASS+}), we see that while  coupling to $k$ Glauber gluons behave like soft gluons.

$l_{1},\ldots, l_{n}$ couple to soft particles $k_{s}$ on the other ends . While coupling to soft particles, Glauber gluons behave like soft gluons as demonstrated in the effective action (\ref{effective action}). In fact the Glauber region can be viewed as the subregion of the soft region in loop momenta integral. Hence couplings between Glauber gluons and $k_{s}$ can be absorbed into those between soft gluons and $k_{s}$ by extending the soft region to include the Glauber region in the loop integrals as one usually does.
In summary Glauber gluons exchanged between active particles and soft particles behave like soft gluons on both ends.

We also notice that couplings between Glauber gluons and ultrasoft particles are power suppressed according to the power counting(\ref{Power coupling}). So are couplings between soft gluons and ultrasoft particles. Hence Glauber gluons behave like soft gluons at leading power of $\lambda$ while coupling to ultrasoft particles or ultrasoft gluons.

In addition propagators of $l_{j}$ behave like those of $q_{j}$ in the special momenta region($|q_{j}^{-}|\ll |q_{j\perp}|, \quad 1\le j\le n$).

According to above facts we see that Glauber gluons exchanged between $k_{s}$ and $k$ behave like soft gluons and can be absorbed into soft Wilson along $+$-direction. For other Glauber gluons exchanged between active and soft particles, we have similar results. In other words,  effects of Glauber gluons exchanged between active and soft particles can be absorbed into zero bins of soft Wilson lines.\footnote{The Wilson lines should be past pointing according to the poles location of $l_{j}^{-}$ in (\ref{eikG+}). One may also see this by simply realizing that final interactions cancel out in $\mathcal{H}(P,\bar{P},q^{+},q^{-})$ as discussed in Sec.\ref{ficanel}}

\subsection{Glauber gluons exchanged between  spectators  and active particles}
\label{asG}

We consider Glauber gluons exchanged between  spectators  and active particles in this subsection. These Glauber gluons are absorbed into collinear Wilson lines according to discussions here.

Without loss of generality, we consider Glauber gluons exchanged between a $-$-collinear spectator $\bar{K}$ and a $+$-collinear active particle $k$($k^{+}>0$ as plus momenta flow from $+$-collinear particles to the hard vertex).  While coupling to spectators, these Glauber gluons behave like collinear gluons $A_{-,\bar{p}}^{\mu}$ with $\bar{p}^{-}=0$. On the other hand, couplings between Glauber and ultrasoft gluons are power suppressed according to the power counting results \ref{Power coupling}. Hence, we can use the eikonal approximation in couplings between Glauber and ultrasoft  gluons without affecting  leading power results. So are couplings between these Glauber gluons and other Glauber gluons. \footnote{Couplings between Glauber and soft gluons are presented in our discussions on Glauber gluons exchanged between soft gluons and collinear particles in Sec.\ref{asG}.}In other words, these Glauber gluons behave like collinear gluons $A_{-,\bar{p}}^{\mu}$ with $\bar{p}^{-}=0$ while coupling to spectators and ultrasoft  and Glauber gluons.

On the other ends, these Glauber gluons couple to $+$-collinear active particles.  According to results in Sec.\ref{CAG}, Such couplings eikonalize at leading power of $\lambda$ and $\eta$ after cancellation of final interactions.  Couplings between these Glauber gluons and $k$ read(\ref{eikG+}
\begin{eqnarray}
\label{GAC+}
&&
\int\frac{\ud^{D} l_{1}}{(2\pi)^{D}}\cdots\int\frac{\ud^{D} l_{n}}{(2\pi)^{D}}
\frac{1}{k^{-}+l_{1}^{-}+\frac{(k_{\perp})^{2}}{2k^{+}}+i\epsilon}\cdots
\frac{1}{k^{-}+l_{1}^{-}+\cdots+l_{n}^{-}+\frac{(k_{\perp})^{2}}{2k^{+}}+i\epsilon}
\nonumber\\
&&\cdots \frac{1}{k^{-}+l_{1}^{-}+\cdots+l_{n}^{-}+\cdots+\frac{(k_{\perp}+\cdots)^{2}}{2k^{+}}+i\epsilon}
\nonumber\\
&&\frac{1}{l_{1\perp}^{2}+i\epsilon}\cdots\frac{1}{l_{n\perp}^{2}+i\epsilon}
\times \text{(terms independent of $l_{j}$)($1\le j\le n$)}
\nonumber\\
&&\times\text{(terms on the other ends of $l_{1},\ldots,l_{n}$)}
\end{eqnarray}
at leading power of $\lambda$ and $\eta$, where $D=4-2\varepsilon$. On the other hand, one may consider couplings between $k$ and $-$-collinear gluons. We denote  momenta of these collinear gluons as $\bar{k}_{1},\ldots,\bar{k}_{n}$. Couplings between these collinear gluons and $k$ can be written as
\begin{eqnarray}
\label{CAC+}
&&\int\frac{\ud^{D} \bar{k}_{1}}{(2\pi)^{D}}\cdots\int\frac{\ud^{D} \bar{k}_{n}}{(2\pi)^{D}}
\frac{1}{k^{-}+\bar{k}_{1}^{-}+\frac{(k_{\perp})^{2}}{2k^{+}}+i\epsilon}\cdots
\frac{1}{k^{-}+\bar{k}_{1}^{-}+\cdots+\bar{k}_{n}^{-}+\frac{(k_{\perp})^{2}}{2k^{+}}+i\epsilon}
\nonumber\\
&&\cdots \frac{1}{k^{-}+\bar{k}_{1}^{-}+\cdots+\bar{k}_{n}^{-}+\cdots+\frac{(k_{\perp}+\cdots)^{2}}{2k^{+}}+i\epsilon}
\nonumber\\
&&\frac{1}{\bar{k}_{1}^{+}\bar{k}_{1}^{-}+\bar{k}_{1\perp}^{2}+i\epsilon}\cdots\frac{1}{\bar{k}_{n}^{+}\bar{k}_{n}^{-}+\bar{k}_{n\perp}^{2}+i\epsilon}
\times \text{(terms independent of $\bar{k}_{j}$)($1\le j\le n$)}
\nonumber\\
&&\times\text{(terms on the other ends of $\bar{k}_{1},\ldots,\bar{k}_{n}$)}
\end{eqnarray}
at leading power of $\lambda$ and $\eta$, where we have made use of the eikonal approximation in couplings between $k$ and $\bar{k}_{1},\ldots,\bar{k}_{n}$.  We see that couplings between $k$ and $l_{1},\ldots l_{n}$ behave like those between $k$ and $\bar{k}_{1},\ldots, \bar{k}_{n}$.


In summary, Glauber gluons exchanged between $\bar{K}$ and $k$ behaves like collinear gluons $A_{-,\bar{p}}^{\mu}$ with $\bar{p}^{-}=0$ on both ends. Propagators of these Glauber gluons behave like those of collinear gluons $A_{-,\bar{p}}^{\mu}$ with $\bar{p}^{-}=0$ too. Hence Glauber gluons exchanged between $\bar{K}$ and $k$ can be absorbed into $-$-collinear Wilson lines by extending the collinear region to include the Glauber region in the loop integrals as one usually does.\footnote{The Wilson lines should be past pointing as final interactions cancel out in $\mathcal{H}(P,\bar{P},q^{+},q^{-})$ as discussed in Sec.\ref{ficanel}}

For general Glauber gluons exchanged between  spectators  and active particles, we have the similar results. In conclusion, Glauber gluons exchanged between  spectators  and active particles can be absorbed into collinear Wilson lines.

\subsection{Glauber gluons exchanged between active particles}
\label{aaG}

We consider effects of Glauber gluons exchanged between  active particles in this subsection. In ladder diagrams of such type, effects of these gluons can be absorbed into soft Wilson lines\cite{Rothstein:2016bsq}. In this subsection, we extend the result to general situation.


Without loss of generality, we consider Glauber gluons($l_{1},\ldots,l_{n}$) exchanged between a $+$-collinear active particle $k$($k^{+}>0$ as plus momenta flow from $+$-collinear particles to the hard vertex) and a $-$-collinear active particle $\bar{k}$($\bar{k}^{-}>0$ as minus momenta flow from $-$-collinear particles to the hard vertex). Couplings between Glauber gluons and active particles eikonalize according to the result in Sec.\ref{CAG}. In other words, we can use the eikonal approximation on both ends of these Glauber gluons.
Couplings between $k$ and Glauber gluons can be written as(\ref{eikG+})
\begin{eqnarray}
\label{AAG+}
&&
\int\frac{\ud^{D} l_{1}}{(2\pi)^{D}}\cdots\int\frac{\ud^{D} l_{n}}{(2\pi)^{D}}
\frac{1}{k^{-}+l_{1}^{-}+\frac{(k_{\perp})^{2}}{2k^{+}}+i\epsilon}\cdots
\frac{1}{k^{-}+l_{1}^{-}+\cdots+l_{n}^{-}+\frac{(k_{\perp})^{2}}{2k^{+}}+i\epsilon}
\nonumber\\
&&\cdots \frac{1}{k^{-}+l_{1}^{-}+\cdots+l_{n}^{-}+\cdots+\frac{(k_{\perp}+\cdots)^{2}}{2k^{+}}+i\epsilon}
\frac{1}{l_{1\perp}^{2}+i\epsilon}\cdots\frac{1}{l_{n\perp}^{2}+i\epsilon}
\nonumber\\
&&
\times \text{(terms independent of $l_{j}$)($1\le j\le n$)}
\nonumber\\
&&\times\text{(terms on the other ends of $l_{1},\ldots,l_{n}$)}
\end{eqnarray}
at leading power of $\lambda$ and $\eta$, where $D=4-2\varepsilon$. One may compare the result with the  cases that $k$ couple to $-$-collinear gluons $\bar{p}_{1},\ldots,\bar{p}_{n}$ or soft gluons $q_{1},\ldots,q_{n}$.   After the eikonal approximation, couplings between $\bar{p}_{1},\ldots,\bar{p}_{n}$ and $k$ read
\begin{eqnarray}
\label{AA-+}
&&\int\frac{\ud^{D} \bar{p}_{1}}{(2\pi)^{D}}\cdots\int\frac{\ud^{D} \bar{p}_{n}}{(2\pi)^{D}}
\frac{1}{k^{-}+\bar{p}_{1}^{-}+\frac{(k_{\perp})^{2}}{2k^{+}}+i\epsilon}\cdots
\frac{1}{k^{-}+\bar{p}_{1}^{-}+\cdots+\bar{p}_{n}^{-}+\frac{(k_{\perp})^{2}}{2k^{+}}+i\epsilon}
\nonumber\\
&&\cdots \frac{1}{k^{-}+\bar{p}_{1}^{-}+\cdots+\bar{p}_{n}^{-}+\cdots+\frac{(k_{\perp}+\cdots)^{2}}{2k^{+}}+i\epsilon}
\nonumber\\
&&\frac{1}{\bar{p}_{1}^{+}\bar{p}_{1}^{-}+\bar{p}_{1\perp}^{2}+i\epsilon}\cdots\frac{1}{\bar{p}_{n}^{+}\bar{p}_{n}^{-}+\bar{p}_{n\perp}^{2}+i\epsilon}
\times \text{(terms independent of $\bar{p}_{j}$)($1\le j\le n$)}
\nonumber\\
&&\times\text{(terms on the other ends of $\bar{p}_{1},\ldots,\bar{p}_{n}$)}
\end{eqnarray}
and couplings between $q_{1},\ldots,q_{n}$ and $k$ read
\begin{eqnarray}
\label{AAS+}
&&\int\frac{\ud^{D} q_{1}}{(2\pi)^{D}}\cdots\int\frac{\ud^{D} q_{n}}{(2\pi)^{D}}
\frac{1}{k^{-}+q_{1}^{-}+\frac{(k_{\perp})^{2}}{2k^{+}}+i\epsilon}\cdots
\frac{1}{k^{-}+q_{1}^{-}+\cdots+q_{n}^{-}+\frac{(k_{\perp})^{2}}{2k^{+}}+i\epsilon}
\nonumber\\
&&\cdots \frac{1}{k^{-}+q_{1}^{-}+\cdots+q_{n}^{-}+\cdots+\frac{(k_{\perp}+\cdots)^{2}}{2k^{+}}+i\epsilon}
\nonumber\\
&&\frac{1}{q_{1}^{+}q_{1}^{-}+q_{1\perp}^{2}+i\epsilon}\cdots\frac{1}{q_{n}^{+}q_{n}^{-}+q_{n\perp}^{2}+i\epsilon}
\times \text{(terms independent of $q_{j}$)($1\le j\le n$)}
\nonumber\\
&&\times\text{(terms on the other ends of $q_{1},\ldots,q_{n}$)}
\end{eqnarray}
at leading power of $\lambda$ and $\eta$.  Comparing (\ref{AAG+}) with (\ref{AA-+}) and (\ref{AAS+}), we see that  Glauber gluons behave like collinear gluons $A_{+,0}^{\mu}$ and soft gluons $A_{s}^{\mu}$ while  coupling to $k$.

For couplings between $l_{1},\ldots,l_{n}$ and $\bar{k}$, we have the similar result and Glauber gluons behave like collinear gluons $A_{-,0}^{\mu}$ and soft gluons $A_{s}^{\mu}$ in these couplings.

According to above discussions, we see that Glauber gluons exchanged between $k$ and $\bar{k}$ behave like soft gluons on both ends.
In addition, propagators of  $l_{1},\ldots,l_{n}$ behave like those of $q_{1},\ldots,q_{n}$ in the special momenta region($|q_{j}^{+}|\ll |q_{j\perp}|$, $|q_{j}^{-}|\ll |q_{j\perp}|$).

For general Glauber gluons exchanged between active particles, we have similar results. Hence effects of these Glauber gluons can be absorbed into soft Wilson lines by extending the soft region to include the Glauber region in loop integralss.\footnote{These Wilson lines should be past pointing as final interactions cancel out in $\mathcal{H}(P,\bar{P},q^{+},q^{-})$ }

\section{Cancellation of spectator-spectator and spectator-soft Glauber exchanges in $\mathcal{H}(P,\bar{P},q^{+},q^{-})$}
\label{canSS}

In this subsection, we prove the cancellation of spectator-spectator and spectator-soft Glauber exchanges in $\mathcal{H}(P,\bar{P},q^{+},q^{-})$. Calculations in \cite{Rothstein:2016bsq} show the cancellation of  spectator-spectator type Glauber exchanges in ladder diagrams. According to our discussions in Sec.\ref{ss0as}, ladder diagrams of Glauber gluons exchanged between spectators in $\mathcal{H}$ can be understood as perturbative series of the object
\begin{eqnarray}
\label{canssla}
&&\sum_{X}<p_{1}^{\prime}\bar{p}_{1}^{\prime}\ldots |U^{\dag}(\infty,-\infty)|X><X|U(\infty,-\infty)|p_{1}\bar{p}_{1}\ldots>
\nonumber\\
&=&
<p_{1}^{\prime}\bar{p}_{1}^{\prime}\ldots |U^{\dag}(\infty,-\infty)U(\infty,-\infty)|p_{1}\bar{p}_{1}\ldots>
\nonumber\\
&=&<p_{1}^{\prime}\bar{p}_{1}^{\prime}\ldots |p_{1}\bar{p}_{1}\ldots>,
\end{eqnarray}
where $|p_{1}\bar{p}_{1}\ldots>$ and $|p_{1}^{\prime}\bar{p}_{1}^{\prime}\ldots>$ represent initial spectators and $|X>$ represents possible final states and $U(t_{2},t_{1})$ represents the time evolution operator of corresponding to spectator-spectator Glauber exchanges. We have made use of the unitarity of time evolution operator in above equation. Eq.(\ref{canssla}) explains the cancellation of ladder diagrams of Glauber gluons exchanged between spectators.

\begin{figure*}
\begin{center}
\includegraphics{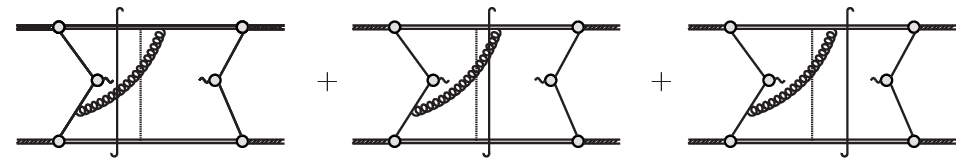}
\end{center}
\caption{An example of diagrams with spectator-active coherence which obstruct the summation over all possible Glauber interactions between spectators\,  where dot lines represent Glauber gluons.}
\label{nonladder}
\end{figure*}

In general cases, spectator-active and soft-active coherence may obstruct the summation over all possible final spectators. For example, one may consider the diagram shown in Fig.\ref{nonladder}.
The gluon exchanged between spectators and active particles should not be collinearin the first two diagrams. Otherwise the two diagrams do not contribute to the process considered here. In other words, the summation over Glauber interactions of spectators is hampered by spectator-active coherence in Fig.\ref{nonladder}.  One should deal with effects of spectator-active and soft-active coherence carefully to get the cancellation of spectator-spectator and spectator-soft Glauber exchanges.

\begin{figure*}
\begin{tabular}{ccc}
\includegraphics[scale=0.3]{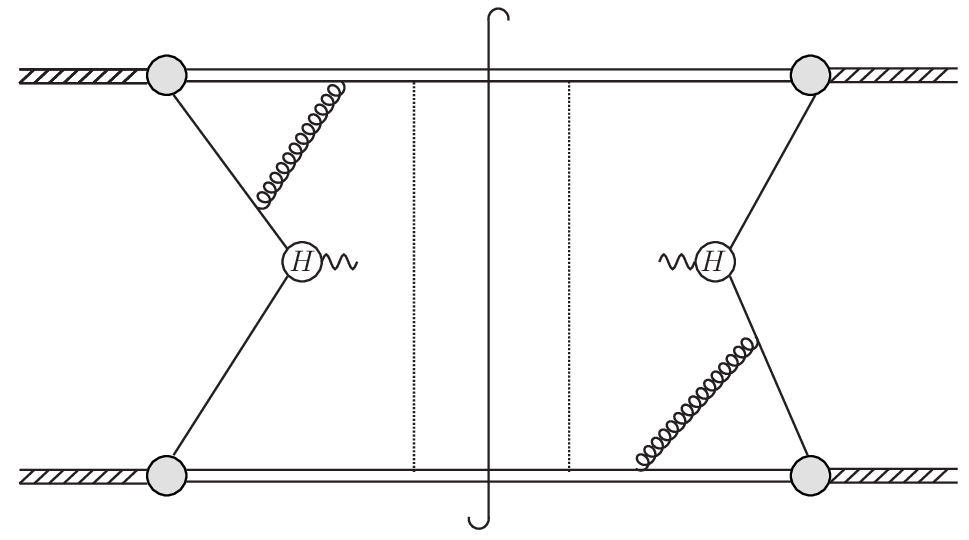}
&
\includegraphics[scale=0.3]{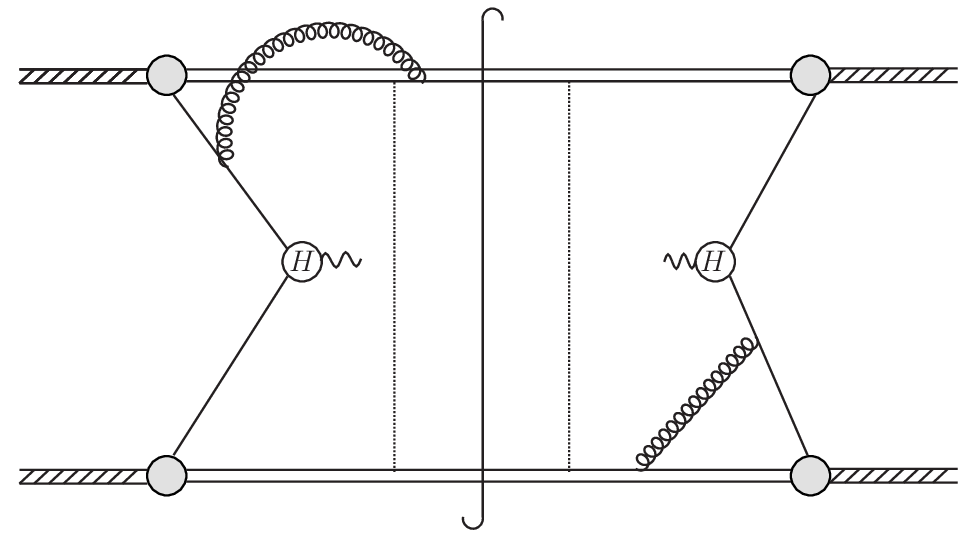}
&
\includegraphics[scale=0.3]{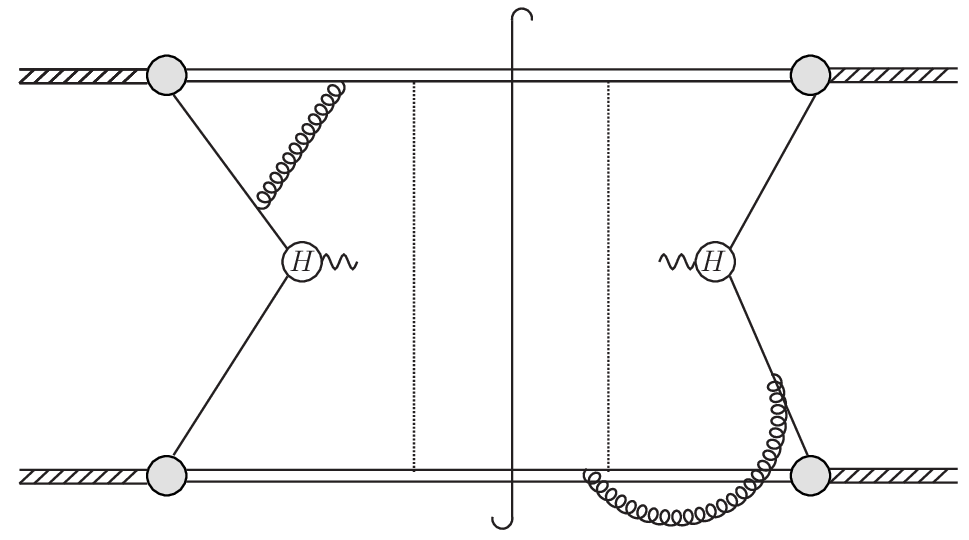}
\\
(a)&(b)&(c)
\end{tabular}
\caption{(a) An example of diagrams with Glauber gluons $A_{+G}$ couple to spectators after non-Glauber couplings.(b)An example of diagrams with Glauber gluons $A_{+G}$ couple to $+$-collinear spectators before non-Glauber couplings. (c)An example of diagrams with Glauber gluons $A_{+G}$ couple to $-$-collinear spectators before non-Glauber couplings. Glauber gluons are represented by dot lines in these diagrams.}
\label{CG+spt}
\end{figure*}

If the Glauber interactions occur after the spectator-active and soft-active interactions like the first diagram in Fig.\ref{CG+spt} then the summation over all states after spectator-active and soft-active interactions is inclusive enough for spectator-spectator and spectator-soft  Glauber exchanges.
The cancellation (\ref{canssla}) can be extended into this case even if there are spectator-active and soft-active interactions. It seems important for us to exclude spectator-active and soft-active interactions after spectator-spectator and spectator-soft  Glauber exchanges like those in Fig.\ref{nonladder}. However, instated of time evolution of collinear and soft states,   we find it convenient to consider evolution of these  states along a nearly light like direction. This is displayed explicitly in following subsections.

According to discussions in Sec.\ref{AG},  Glauber gluons coupling to active particles should be absorbed into zero bins of collinear or soft gluons.  Hence we view Glauber interactions of active particles as collinear or soft interactions of these particles. While referring to Glauber interactions, we always mean spectator-spectator and spectator-soft  Glauber exchanges in following texts.

\subsection{$\widetilde{n_{+}}\cdot x $-evolution in $\mathcal{H}(P,\bar{P},q^{+},q^{-})$}
\label{x+H}

It is convenient to consider the $x^{+}$(or $x^{-}$)-evolution instead of time evolution of states  in $\mathcal{H}(P,\bar{P},q^{+},q^{-})$. Such evolution is crucial in proof of factorization theorem of Drell-Yan process in the frame of perturbative QCD\cite{CSS:1988}. In this subsection, we consider a time-like evolution of collinear and soft states which is approximately equivalent to $x^{+}$ or $x^{-}$-evolution  in the frame of effective theory.

Let us bring in a time like directions $\widetilde{n_{+}}$ and a space like direction $\widehat{n_{+}}$  at first
\begin{eqnarray}
\widetilde{n_{+}}^{\mu}&\equiv&(\widetilde{n_{+}}^{+},\widetilde{n_{+}}^{-},\widetilde{n_{+}}_{\perp}^{\mu})
\equiv\frac{1}{\sqrt{2}}(\frac{1}{\sqrt{\lambda\omega}},\sqrt{\lambda\omega},0)
\nonumber\\
\widehat{n_{+}}^{\mu}&\equiv&(\widehat{n_{+}}^{+},\widehat{n_{+}}^{-},\widetilde{n_{+}}_{\perp}^{\mu})
\equiv\frac{1}{\sqrt{2}}(\frac{1}{\sqrt{\lambda\omega}},-\sqrt{\lambda\omega},0),
\end{eqnarray}
where $\omega$ is a positive constant of order $1$. We have
\begin{eqnarray}
\widetilde{n_{+}}^{2}=1, && \frac{\widetilde{n_{+}}^{-}}{\widetilde{n_{+}}^{+}}=\lambda\omega
\nonumber\\
\widehat{n_{+}}^{2}=-1, && \frac{\widehat{n_{+}}^{-}}{\widehat{n_{+}}^{+}}=-\lambda\omega.
\end{eqnarray}
That is to say $\widetilde{n_{+}}$($\widehat{n_{+}}$) is time(space)-like and nearly parallel to the plus direction.

One can choose a reference system so that
\begin{equation}
\widetilde{n_{+}}^{\mu}\equiv(\widetilde{n_{+}}^{0},\vec{\widetilde{n_{+}}})
\to (1,\vec{0})
\end{equation}
in the new reference system as $\widetilde{n_{+}}$ is time like. According to similar skills in Sec.\ref{ficanel} one has
\begin{eqnarray}
&&\sum_{X}\int\ud^{4} x_{1}\int\ud^{4} x_{2}\int\ud^{4} x_{3}\int\ud^{4} x_{4}\int\ud^{4}z e^{iq\cdot z}
\nonumber\\
&&\big<H_{1}H_{2}|\bar{T}\{\mathcal{O}^{\dag}(x_{1})\bar{\mathcal{O}}^{\dag}(x_{2})J^{\dag}(z)U^{\dag}(\infty,-\infty)\}|H(q)X\big>
\nonumber\\
&&\big<H(q)X|T\{\mathcal{O}(x_{3})\bar{\mathcal{O}}(x_{4})J(0)U(\infty,-\infty)\}|H_{1}H_{2}\big>
\nonumber\\
&=&\sum_{X}\int\ud^{4} x_{1}\int\ud^{4} x_{2}\int\ud^{4} x_{3}\int\ud^{4} x_{4}\int\ud^{4}z e^{iq\cdot z}
\nonumber\\
&&\big<H_{1}H_{2}|\mathcal{O}^{\dag}(x_{1})\bar{\mathcal{O}}^{\dag}(x_{2})\bar{T}\{J^{\dag}(z)U^{\dag}(\max\{\widetilde{n_{+}}\cdot z,0\},-\infty)\}
|H(q)X\big>
\nonumber\\
&&\big<H(q)X|T\{J(0)U(\max\{\widetilde{n_{+}}\cdot z,0\},-\infty)\}\mathcal{O}(x_{3})\bar{\mathcal{O}}(x_{4})|H_{1}H_{2}\big>
\end{eqnarray}
in the new reference system.

According to the Lorentz invariance of QCD, one has
\begin{eqnarray}
\label{canH+}
&&\mathcal{H}(P,\bar{P},q^{+},q^{-})
\nonumber\\
&\equiv&\int\frac{\ud^{2}q_{\perp}}{(2\pi)^{2}} \sum_{X}\int\ud^{4} x_{1}\int\ud^{4} x_{2}\int\ud^{4} x_{3}\int\ud^{4} x_{4}\int\ud^{4}z e^{iq\cdot z}
\nonumber\\
&&\big<P\bar{P}|\mathcal{O}^{\dag}(x_{1})\bar{\mathcal{O}}^{\dag}(x_{2})\bar{T}\{J^{\dag}(z)U_{\widetilde{n_{+}}}^{\dag}(\max\{\widetilde{n_{+}}\cdot z,0\},-\infty)\}
|H(q)X\big>
\nonumber\\
&&\big<H(q)X|T\{J(0)U_{\widetilde{n_{+}}}(\max\{\widetilde{n_{+}}\cdot z,0\},-\infty))\}\mathcal{O}(x_{3})\bar{\mathcal{O}}(x_{4})|P\bar{P}\big>
\end{eqnarray}
in the center of mass frame of initial hadrons, where $U_{\widetilde{n_{+}}}$ represents the evolution operator along the $\widetilde{n_{+}}\cdot x$ direction. We do not concern detail of $U_{\widetilde{n_{+}}}$ here.   That is to say, interactions of which the coordinates $\widetilde{n_{+}}\cdot x$ are greater than that of the hard collision cancel out in $\mathcal{H}(P,\bar{P},q^{+},q^{-})$. \footnote{One may try to consider evolution of states along the plus or minus direction directly in $\mathcal{H}(P,\bar{P},q^{+},q^{-})$. However, the boundary condition in the limit $x^{+}\to\infty$ or $x^{-}\to\infty$ is not clear for us.}

\subsection{Couplings involving Glauber gluons $A_{+G}$ in $\mathcal{H}(P,\bar{P},q^{+},q^{-})$}
\label{CA+G}

In this subsection we consider Glauber gluons  $A_{+G}$ exchanged between $+$-collinear and other particles. We prove the cancellation of  Glauber gluons $A_{+G}$ coupling to collinear or soft particles at vertexes of which the coordinates $\widetilde{n_{+}}\cdot x$ are smaller than those of some vertexes free from $A_{+G}$(except for the hard vertex).

To distinguish vertexes involving $A_{+G}$ and those free from $A_{+G}$, we denote coordinates of couplings between $A_{+G}$ and $+$-collinear particles as $y_{i}$ and those between $A_{+G}$ and $-$-collinear and soft particles as $y_{i}^{\prime}$ and those free from  $A_{+G}$ as $z_{i}$($i=1,\ldots$). Without loss of generality , we consider a  Glauber gluon $l_{1}$ exchanged between the two vertexes $y_{1}$ and $y_{1}^{\prime}$. The propagator of $l_{1}$ is independent of $l_{1}^{+}$ and $l_{1}^{-}$ at leading power of $\lambda$ and $\eta$. In addition $l_{1}^{+}$($l_{1}^{-}$) can be neglected at the vertex $y_{1}$ ($y_{1}^{\prime}$). \footnote{$l_{1}^{+}\lesssim O(Q\lambda)\ll p^{+}\sim O(Q)$ for $+$-collinear particles and  $l_{1}^{-}\sim O(Q\lambda^{2})\ll l_{s}^{-}\sim O(Q\lambda) \ll \bar{p}^{-}\sim O(Q)$ where $p^{+}$ represents plus momenta of $+$-collinear particles and $l_{s}^{-}$ and $\bar{p}^{-}$ represent minus momenta of soft and $-$-collinear particles.}

Considering that the propagator of $l_{1}$ is independent of $l_{1}^{+}$ and $l_{1}^{-}$ at leading power of $\lambda$ and $\eta$. As a result, $\mathcal{H}(P,\bar{P},q^{+},q^{-})$ relies on $l_{1}^{+}$ and $l_{1}^{-}$ only through the term
\begin{equation}
 \exp\left(-il_{1}^{-}y_{1}^{+}+il_{1}^{+}y_{1}^{\prime -}\right)
 \end{equation}
and integrals over $l_{1}^{-}$ and $l_{1}^{+}$ read
\begin{eqnarray}
\int\frac{\ud l_{1}^{-}}{2\pi }\frac{\ud l_{1}^{+}}{2\pi} \exp\left(-il_{1}^{-}y_{1}^{+}+il_{1}^{+}y_{1}^{\prime -}\right)&=&\delta(y_{1}^{+})\delta(y_{1}^{\prime -})
\end{eqnarray}

The vertex $y_{1}$ and the hard vertex are connected through $+$-collinear particles. For a $+$-collinear particle with momenta $\xi P+k$($\xi\sim O(1)$, $k^{\mu}\lesssim Q\lambda$ for $+$-collinear external lines),  the propagator reads
\begin{equation}
\int\frac{\ud^{4}k}{(2\pi)^{4}}\frac{N(P)}{2\xi P^{+}k^{-}+(k_{\perp})^{2}+i\epsilon}e^{-ik\cdot (x_{1}-x_{2})}
\propto \delta(x_{1}^{-}-x_{2}^{-}),
\end{equation}
where  $N(P)$ represents possible numerators in propagators of collinear particles.
We have
\begin{equation}
 y_{1}^{-}=0,\quad \delta(y_{1}^{+})=\sqrt{\frac{\lambda\omega}{2}}\delta\left(\widetilde{n_{+}}\cdot y_{1}\right)
\end{equation}
on left of the final cut and
\begin{equation}
 y_{1}^{-}=z^{-},\quad \delta(y_{1}^{+})\simeq\delta(y_{1}^{+}-z^{+})
 = \sqrt{\frac{\lambda\omega}{2}}\delta\left(\widetilde{n_{+}}\cdot y_{1}-\widetilde{n_{+}}\cdot z\right)
\end{equation}
on right of the final cut, where we have made use of the fact
\begin{equation}
z^{+}\sim z^{-}\sim 1/Q.
\end{equation}

For the vertex $y_{1}^{\prime}$, we have
\begin{eqnarray}
 \delta(y_{1}^{\prime -})\simeq\frac{1}{\sqrt{2\lambda \omega}} \delta\left(\widetilde{n_{+}}\cdot y_{1}^{\prime}\right)
\end{eqnarray}
on left of the final cut and
\begin{equation}
\delta(y_{1}^{\prime -})\simeq \delta(y_{1}^{\prime -}-z_{1}^{-}) \simeq\frac{1}{\sqrt{2\lambda \omega}} \delta\left(\widetilde{n_{+}}\cdot y_{1}^{\prime}-\widetilde{n_{+}}\cdot y_{1}\right)
\end{equation}
on right of the final cut.
Hence
\begin{equation}
 \widetilde{n_{+}}\cdot z_{i}\le 0\Rightarrow  \widetilde{n_{+}}\cdot z_{i}\le \widetilde{n_{+}}\cdot y_{1},
 \quad \widetilde{n_{+}}\cdot z_{i}\le \widetilde{n_{+}}\cdot y_{1}^{\prime}
 \end{equation}
on left of the final cut and
 \begin{equation}
 \widetilde{n_{+}}\cdot z_{i}\le \widetilde{n_{+}}\cdot z\Rightarrow  \widetilde{n_{+}}\cdot z_{i}\le \widetilde{n_{+}}\cdot y_{1},
 \quad \widetilde{n_{+}}\cdot z_{i}\le \widetilde{n_{+}}\cdot y_{1}^{\prime}
 \end{equation}
on right of the final cut.  That is to say
\begin{equation}
\widetilde{n_{+}}\cdot z_{i}\le \widetilde{n_{+}}\cdot y_{1},\quad \widetilde{n_{+}}\cdot z_{i}\le \widetilde{n_{+}}\cdot y_{1}^{\prime}
\end{equation}
on both sides of the final cut.

For other vertexes involving Glauber gluons $A_{+G}$  we have similar results.  We conclude that Glauber gluons $A_{+G}$ should couple to collinear or soft particles at vertexes with coordinates $\widetilde{n_{+}}\cdot y_{i}$ greater than those of vertexes free from Glauber gluons $A_{+G}$(except for the hard vertex), otherwise effects of  Glauber gluons $A_{+G}$ cancel out at leading power of $\lambda$ and $\eta$.

\subsection{Cancellation of Glauber gluon $A_{+G}$}
\label{canA+G}

In this subsection, we prove the cancellation of Glauber gluons $A_{+G}$. These Glauber gluons are exchanged between $+$- and $-$- collinear spectators or between $+$-collinear spectator  and soft particles.

According to results in Sec.\ref{CA+G}, Glauber gluons  $A_{+G}$ should couple to collinear and soft particles at vertexes with the coordinates $\widetilde{n_{+}}\cdot x$ greater than those of vertexes free from $A_{+G}$(except for vertexes in the hard subprocess).
After absorption of Glauber gluons coupling to active particles into collinear and Wilson lines, we have
\begin{eqnarray}
&&\mathcal{H}(P,\bar{P},q^{+},q^{-})
\nonumber\\
&\equiv&\int\frac{\ud^{2}q_{\perp}}{(2\pi)^{2}} \sum_{X}\int\ud^{4} z e^{iq\cdot z}
\big<P\bar{P}|\bar{T}{\mathcal{O}^{\dag}(S_{+},S_{-},W_{+},W_{-})(x)}
|X\big>
\nonumber\\
&&\big<X|T\{\mathcal{O}(S_{+},S_{-},W_{+},W_{-})(0)\}|P\bar{P}\big>\otimes h(z),
\end{eqnarray}
where $h(z)$ represents contributions of the hard scattering subprocess and    $\mathcal{O}(S_{+},S_{-},W_{+},W_{-})$ represents Wilson line structure of the hard vertex and
\begin{eqnarray}
W_{+}(x)&=&P\exp(ig\int_{-\infty}^{0}\ud s  A_{+}^{\phantom {+}+}(0,x^{-}+s,0)
\nonumber\\
S_{+}(x)&=&P\exp(ig\int_{-\infty}^{0}\ud s\cdot A_{s}^{\phantom {s}-}(x^{+}+s,0,0)
\nonumber\\
W_{-}(x)&=&P\exp(ig\int_{-\infty}^{0}\ud s\cdot A_{-}^{\phantom {-}-}(x^{+}+s,0,0)
\nonumber\\
S_{-}(x)&=&P\exp(ig\int_{-\infty}^{0}\ud s\cdot A_{s}^{\phantom {s}+}(0,x^{-}+s,0).
\end{eqnarray}
$\widetilde{n_{+}}\cdot x$ evolution of fields in these Wilson lines is determined by the operator
\begin{equation}
\label{evolution+}
U_{SCET_{G}}(\widetilde{n_{+}}\cdot x_{1},\widetilde{n_{+}}\cdot x_{2}),
\end{equation}
where $U_{SCET_{G}}$ represents the $n\cdot x$ evolution operator related to the action (\ref{effective action}) and
\begin{equation}
A^{\mu}(x)=U_{SCET_{G}}(\widetilde{n_{+}}\cdot x,0)A^{\mu}(0,\widehat{n_{+}}\cdot x,x_{\perp})U_{SCET_{G}}^{\phantom{SCET_{G}}\dag}(\widetilde{n_{+}}\cdot x,0)
\end{equation}
with $A^{\mu}=A_{+}^{\phantom{+}\mu},A_{-}^{\phantom{-}\mu},A_{s}^{\mu}$.

According to results in Sec.\ref{CA+G}, Glauber gluons $A_{+G}$ couple to spectators and soft particles at vertexes with the coordinates $n\cdot x$ greater than those of free form $A_{+G}$ in the matrix elements
\begin{equation}
\big<X|T\{\mathcal{O}(S_{+},S_{-},W_{+},W_{-})(0)\}|P\bar{P}\big>
\end{equation}
and
\begin{equation}
\big<P\bar{P}|\bar{T}{\mathcal{O}^{\dag}(S_{+},S_{-},W_{+},W_{-})(x)}|X\big>.
\end{equation}
We consider all perturbative order interactions involving $A_{+G}$ and make the perturbative expansion about interactions free from  $A_{+G}$. We have
\begin{eqnarray}
&&\mathcal{H}(P,\bar{P},q^{+},q^{-})
\nonumber\\
&=&\sum_{m_{1},m_{2}}
\int\frac{\ud^{2}q_{\perp}}{(2\pi)^{2}} \sum_{X}\int\ud^{4} z
\int_{-\infty}^{\widetilde{n_{+}}\cdot z} \ud s_{1}
\int_{-\infty}^{0} \ud s_{2}
e^{iq\cdot z}
\nonumber\\
&&\big<P\bar{P}|\mathcal{O}_{m_{1}}^{\dag}(s_{1})U_{+G}^{\phantom{+G}\dag}(\widetilde{n_{+}}\cdot z,s_{1})
|X\big>
\nonumber\\
&&\big<X|U_{+G}(0,s_{2})\mathcal{O}_{m_{2}}(s_{2})|P\bar{P}\big>\otimes h(z),
\end{eqnarray}
where $\mathcal{O}_{m_{1}}^{\dag}$ and $\mathcal{O}_{m_{2}}$ represent $m_{1}$-th and $m_{2}$-th perturbative order interactions free from $A_{+G}$ and $U_{+G}$ represents the $\widetilde{n_{+}}\cdot x$ evolution operator corresponding to the spectator-spectator and spectator-soft  Glauber exchanges.

Considering that
\begin{equation}
\widetilde{n_{+}}\cdot x\simeq 0,\quad \widetilde{n_{+}}\cdot z
\end{equation}
for interactions involving Glauber gluons $A_{+G}$\footnote{While considering next leading power results, one has $\widetilde{n_{+}}\cdot x\sim O(\lambda)$ and  $\widetilde{n_{+}}\cdot x<0,\quad\widetilde{n_{+}}\cdot z$} as discussed in Sec.\ref{CA+G}, we have
\begin{equation}
U_{+G}(0,s_{2})\simeq U_{+G}(\infty,-\infty)
\end{equation}
on left of the final cut and
\begin{equation}
U_{+G}^{\phantom{+G}\dag}(\widetilde{n_{+}}\cdot z,s_{1})\simeq U_{+G}^{\phantom{+G}\dag}(\infty,-\infty)
\end{equation}
on right of the final cut.  Hence
\begin{eqnarray}
\label{canHA+G}
&&\mathcal{H}(P,\bar{P},q^{+},q^{-})
\nonumber\\
&\simeq&\sum_{m_{1},m_{2}}
\int\frac{\ud^{2}q_{\perp}}{(2\pi)^{2}} \sum_{X}\int\ud^{4} z
\int_{-\infty}^{\widetilde{n_{+}}\cdot z} \ud s_{1}
\int_{-\infty}^{0} \ud s_{2}
e^{iq\cdot z}
\nonumber\\
&&\big<P\bar{P}|\mathcal{O}_{m_{1}}^{\dag}(s_{1})U_{+G}^{\phantom{+G}\dag}(\infty,-\infty)
|X\big>
\nonumber\\
&&\big<X|U_{+G}(\infty,-\infty)\mathcal{O}_{m_{2}}(s_{2})|P\bar{P}\big>\otimes h(z)
\nonumber\\
&=&\sum_{m_{1},m_{2}}
\int\frac{\ud^{2}q_{\perp}}{(2\pi)^{2}} \sum_{X}\int\ud^{4} z
\int_{-\infty}^{\widetilde{n_{+}}\cdot z} \ud s_{1}
\int_{-\infty}^{0} \ud s_{2}
e^{iq\cdot z}
\nonumber\\
&&\big<P\bar{P}|\mathcal{O}_{m_{1}}^{\dag}(s_{1})
|X\big>
\big<X|\mathcal{O}_{m_{2}}(s_{2})|P\bar{P}\big>\otimes h(z)
\end{eqnarray}
%
at leading power of $\lambda$ and $\eta$.
That is, spectator-spectator and spectator-soft type Glauber exchanges of $A_{+G}$ cancel out in $\mathcal{H}(P,\bar{P},q^{+},q^{-})$.

\subsection{Cancellation of Glauber gluons $A_{-G}$}
\label{canA-G}

In this subsection, we consider Glauber gluons $A_{-G}$ and outline the proof of cancellation of $A_{-G}$.

According to results in Sec.\ref{AG} and Sec.\ref{canA+G}, contributions of Glauber gluons $A_{+G}$ can be absorbed into those of collinear gluons $A_{+G}$ and soft gluons $A_{s}$ by choosing suitable collinear and soft Wilson lines. Hence We can drop the Glauber gluons $A_{+G}$  in calculations of $\mathcal{H}(P,\bar{P},q^{+},q^{-})$. After this we consider the  $\widetilde{n_{-}}\cdot x$-evolution, where the direction $\widetilde{n_{-}}^{\mu}$ is defined as
\begin{equation}
\widetilde{n_{-}}^{\mu}\equiv(\widetilde{n_{-}}^{+},\widetilde{n_{-}}^{-},\widetilde{n_{-}}_{\perp}^{\mu})
\equiv\frac{1}{\sqrt{2}}(\sqrt{\lambda\omega},\frac{1}{\sqrt{\lambda\omega}},0),
\end{equation}
with $\omega$ a positive constant of order $1$. Attributing to proofs similar to those in Sec.\ref{x+H}, interactions with the coordinates $\widetilde{n_{-}}\cdot x$ greater than those of the hard collision cancel out in $\mathcal{H}(P,\bar{P},q^{+},q^{-})$. We have
\begin{eqnarray}
\label{canH-}
&&\mathcal{H}(P,\bar{P},q^{+},q^{-})
\nonumber\\
&\equiv&\int\frac{\ud^{2}q_{\perp}}{(2\pi)^{2}} \sum_{X}\int\ud^{4} x_{1}\int\ud^{4} x_{2}\int\ud^{4} x_{3}\int\ud^{4} x_{4}\int\ud^{4}z e^{iq\cdot z}
\nonumber\\
&&\big<P\bar{P}|\mathcal{O}^{\dag}(x_{1})\bar{\mathcal{O}}^{\dag}(x_{2})\bar{T}\{J^{\dag}(z)U_{\widetilde{n_{-}}}^{\dag}(\max\{\widetilde{n_{-}}\cdot z,0\},-\infty)\}
|H(q)X\big>
\nonumber\\
&&\big<H(q)X|T\{J(0)U_{\widetilde{n_{-}}}(\max\{\widetilde{n_{-}}\cdot z,0\},-\infty))\}\mathcal{O}(x_{3})\bar{\mathcal{O}}(x_{4})|P\bar{P}\big>
\end{eqnarray}
in the center of mass system of initial hadrons, where $U_{\widetilde{n_{-}}}$ represents the evolution operator along the $\widetilde{n_{-}}\cdot x$ direction with the Glauber gluons $A_{+G}$ removed from the effective action.

We then repeat the discussions in Sec.\ref{CA+G}  and see that effects of $A_{-G}$ cancel out at leading power of $\lambda$  and $\eta$ if  Glauber gluons $A_{-G}$ couple to collinear or soft particles at vertexes of which the coordinates $\widetilde{n_{-}}\cdot x$ are smaller than those of some vertexes free from $A_{+G}$(except for vertexes in the hard subprocess).

According to proofs similar to those in  Sec.\ref{canA+G}, we conclude that spectator-spectator and spectator-soft type Glauber gluons(both $A_{+G}$ and $A_{-G}$)  cancel out in $\mathcal{H}(P,\bar{P},q^{+},q^{-})$.

In conclusion, spectator-spectator and spectator-soft Glauber exchanges  cancel out in $\mathcal{H}(P,\bar{P},q^{+},q^{-})$ and other Glauber gluons are equivalent to zero bins of collinear and soft gluons according to discussions in Sec.\ref{AG} and  Sec.\ref{canSS}.

\section{Graphic cancellation of Glauber gluons}
\label{cangrpah}

For readers who are familiar with graphic cancellation of Glauber gluons\cite{B:1985,CSS:1985,CSS:1988} rather than operator level skills in this paper,  we discuss graphic aspects of the cancellation of Glauber gluons in this section. One should not view discussions in this section as strict proof of graphic cancellation of Glauber gluons. Instead we prefer to explain how graphic cancellation of Glauber gluons is related to the operator level cancellation in Sec.\ref{canSS}.

The key point is the unitarity of the $\widetilde{n_{+}}\cdot x$ and $\widetilde{n_{-}}\cdot x$-evolution of collinear and soft particles. Although $\widetilde{n_{+}}$ and $\widetilde{n_{-}}$ are time like, they are approximately equivalent to the $+$ and $-$ direction(with correction of order $\lambda$) in the center of mass frame of initial hadrons. Hence it is not surprise that the $\widetilde{n_{+}}\cdot x$ and $\widetilde{n_{-}}\cdot x$-evolution of collinear and soft particles are approximately equivalent to $x^{+}$ or $x^{-}$-evolution of them.

\subsection{$\widetilde{n_{+}}\cdot x$ evolution of $+$-collinear particles}
\label{x+evo+}


Let us start from couplings involving $+$-collinear particles  at an arbitrary point $y$.
We denote momenta of particles connecting to the vertex $y$ as $l_{i}$($i=1,\ldots$), which are defended as flow into the vertex $y$. Interactions with the coordinates $\widetilde{n_{+}}\cdot x$ greater than those of the hard collision cancel out in $\mathcal{H}(P,\bar{P},q^{+},q^{-})$. Hence
\begin{equation}
\widetilde{n_{+}}\cdot y\le \max\{\widetilde{n_{+}}\cdot z,0\},
\end{equation}
where $z$ and $0$ are the coordinates of the hard collision. At the vertex $y$, one has
 \begin{eqnarray}
&&\int \ud^{4}y \exp\left(-i\sum l_{i}\cdot y\right )\theta\left (\max\{\widetilde{n_{+}}\cdot z,0\}-\widetilde{n_{+}}\cdot y\right )
\nonumber\\
&=&(2\pi)^{3}
\delta\left(\widehat{n_{+}} \cdot\sum l_{i}\right)
\delta^{(2)}\left(\sum l_{i\perp}\right)
\nonumber\\
&&
\frac{i\exp\left(\left(\sum \widetilde{n_{+}}\cdot l_{i}\right)\max\{\widetilde{n_{+}}\cdot z,0\}\right)}{\sum\widetilde{n_{+}}\cdot l_{i}+i\epsilon}
.
\end{eqnarray}
The term
\begin{equation}
\exp\left(\left(\sum \widetilde{n_{+}}\cdot l_{i}\right)\max\{\widetilde{n_{+}}\cdot z,0\}\right)
\end{equation}
contributes to the momenta conservation $\delta$-function of the hard subprocess and can be dropped here.

Without loss of generality, we define $l_{1}$ as the momentum of a $+$-collinear internal line connecting to the vertex $y$ and have
\begin{eqnarray}
&&\widehat{n_{+}}\cdot l_{1}=-\sum_{i\ne 1} \widehat{n_{+}}\cdot l_{i}
\nonumber\\
&\Rightarrow&l_{1}^{+}=-\sum_{i\ne 1} l_{i}^{+}+\frac{1}{\lambda\omega}\sum l_{i}^{-}.
\end{eqnarray}
Considering that
\begin{equation}
l_{i}^{-}\sim O(Q\lambda^{2})
\end{equation}
for $+$-collinear and Glauber gluons $A_{+G}$.\footnote{Couplings between collinear and soft(ultrasoft) particles eikonalize. The coordinates $x^{-}$ of $+$-collinear particles remain unchanged in these couplings. In accordance with our discussions for couplings between Glauber gluons here,  $\widetilde{n_{+}}\cdot x$ and $x^{+}$-order  of $+$-collinear states are  equivalent to each other even if one consider soft(ultrasoft) interactions of these states.} We have
\begin{equation}
l_{1}^{+}\simeq -\sum_{i\ne 1} l_{i}^{+}
\Rightarrow \delta\left(\widehat{n_{+}} \cdot\sum l_{i}\right)\simeq \sqrt{\frac{2}{\lambda\omega}}\delta\left(\sum l_{i}^{+}\right)
\end{equation}
\begin{equation}
\label{app+n+}
\frac{i\delta\left(\widehat{n_{+}} \cdot\sum l_{i}\right)}{\sum\widetilde{n_{+}}\cdot l_{i}+i\epsilon}
\simeq
\frac{i\delta\left(\sum l_{i}^{+}\right)}{\sum l_{i}^{-}+i\epsilon}
\end{equation}
That is to say, couplings involving a $+$-collinear particle $l_{1}$(except for hard interactions) can be calculated through the Feynman diagram skill except for that one should make the substitution
\begin{equation}
\label{substitutioncon+}
(2\pi)^{4}\delta^{(4)}(l_{1}^{\mu}+\ldots)\to  (2\pi)^{3}\delta(l_{1}^{+}+\ldots)\delta^{(2)}(l_{1\perp}+\ldots)
\frac{i}{l_{1}^{-}+\cdots+i\epsilon}
\end{equation}
for the $\delta$-function corresponding to momenta conservation.

We then integral out minus momenta of $+$-collinear internal lines. For an arbitrary $+$-collinear internal line $k$,  $\mathcal{H}$ relies on $l^{-}$ through the propagator
\begin{equation}
\frac{N(l^{+})}{2l^{+}l^{-}+(l_{\perp})^{2}+i\epsilon}.
\end{equation}
 and the two vertexes
\begin{equation}
\frac{1}{-l^{-}+\ldots+q_{i}^{-}+\ldots+i\epsilon},\quad\frac{1}{l^{-}+\ldots+q_{j}^{-}+\ldots+i\epsilon},
\end{equation}
where $q_{i}$ and $q_{j}$ represent momenta of Glauber and soft(ultrasoft) gluons. We take $l^{+}>0$ as plus momenta of $+$-collinear particle flow from the the initial particle $P$ to the hard vertex or final cut.  We can then integrate out $l^{-}$ by taking the residue of the pole locating in the upper half plane. After this operation, we have
\begin{eqnarray}
&&\frac{1}{-l^{-}+\ldots+q_{i}^{-}+\ldots+i\epsilon}\frac{1}{l^{-}+\ldots+q_{j}^{-}+\ldots+i\epsilon}\frac{N(l^{+})}{2l^{+}l^{-}+(l_{\perp})^{2}+i\epsilon}
\nonumber\\
&\to& \frac{-\pi iN(l^{+})}{l^{+}}\frac{1}{q_{i}^{-}+\ldots+q_{j}^{-}+\ldots+i\epsilon}\frac{1}{q_{i}^{-}+\ldots+\frac{(l_{\perp})^{2}}{2l^{+}}+i\epsilon},
\end{eqnarray}
We repeat this procedure and get terms shaped like
\begin{equation}
\label{fincL+}
\frac{1}{P^{-}+\sum q_{i}^{-}+\sum\frac{(k_{i\perp})^{2}}{2k_{i}^{+}}+\sum\frac{(l_{i\perp})^{2}}{2l_{i}^{+}}+i\epsilon}
\end{equation}
on left of the final cut and
\begin{equation}
\label{fincR+}
\frac{1}{P^{-}+\sum q_{j}^{-}+\sum\frac{(k_{j\perp})^{2}}{2k_{j}^{+}}+\sum\frac{(l_{j\perp})^{2}}{2l_{j}^{+}}-i\epsilon}
\end{equation}
on right of the final cut, where $q_{i}$($q_{j}$) represent momenta of Glauber and soft(ultrasoft) gluons and $l_{i}$($l_{j}$) represent momenta of $+$-collinear internal lines and $k_{i}$($k_{j}$) represent momenta of on-shell $+$-collinear particles. The summation of $l_{i}$ ($l_{j}$) is made over states with the coordinates $\widetilde{n_{+}}\cdot x$ between two given vertexes.    The summation of $q_{i}$($q_{j}$) and $k_{i}$($k_{j}$)  are made over states connecting to vertexes with the coordinates $\widetilde{n_{+}}\cdot x$ smaller than those of $l_{i}$($l_{j}$). We should mention that $l_{i}$($l_{j}$) may be internal lines and $l_{i\perp}$($l_{j\perp}$) may relies on transverse momenta of external lines and Glauber and soft(ultrasoft) gluons through the $\delta$-function of transverse momenta conservation.

Compared with the light-cone perturbative series in \cite{CSS:1988},  terms of the type (\ref{fincL+}) and (\ref{fincR+}) correspond to contributions of states with the coordinates $x^{+}$(or the coordinates $\widetilde{n_{+}}\cdot x$ according to the approximation (\ref{app+n+}))  smaller than that of the hard collision. Interactions with the coordinates $\widetilde{n_{+}}\cdot x$  greater than that of the hard collision cancel out in $\mathcal{H}(P,\bar{P},q^{+},q^{-})$ as demonstrated in Sec.\ref{x+H}.

\subsection{$\widetilde{n_{+}}\cdot x$ evolution of $-$-collinear and soft particles}
\label{x+evo-S}

We then consider couplings free from $+$-collinear particles.
Considering that
\begin{equation}
l_{i}^{\mu}\sim O(Q\lambda), \quad l_{i}^{\mu}\sim O(Q\lambda^{2})
\end{equation}
for soft and ultrasoft particles and
\begin{equation}
l_{i}^{\mu}\lesssim O(Q\lambda)
\end{equation}
for Glauber gluons, we have\footnote{Couplings between soft and ultrasoft particles are power suppressed and can neglected. So are couplings between Glauber gluons and ultrasoft particles and those between Glauber gluons.}
\begin{eqnarray}
&&\frac{i\delta\left(\widehat{n_{+}} \cdot\sum l_{i}\right)}{\sum\widetilde{n_{+}}\cdot l_{i}+i\epsilon}
\nonumber\\
&\simeq&\frac{1}{\sqrt{2\lambda\omega}}\frac{i\delta\left (\frac{1}{\sqrt{2\lambda\omega}}\sum l_{i}^{-}\right )}{ \sum l_{i}^{+}+i\epsilon}
\nonumber\\
&=& \frac{i\delta\left(\sum l_{i}^{-}\right)}{ \sum l_{i}^{+}+i\epsilon}.
\end{eqnarray}
Hence we should make the substitution
\begin{equation}
\label{substitutiongs}
(2\pi)^{4}\delta^{(4)}\left(\sum l_{i}^{\mu}\right )\to  (2\pi)^{3}\delta\left (\sum l_{i}^{-}\right )\delta^{(2)}\left (\sum l_{i\perp}\right )
\frac{i}{\sum l_{i}^{+}+i\epsilon}
\end{equation}
in vertexes free from $+$-collinear particles.

We then integral out plus momenta of $-$-these particles. For an arbitrary  internal line $\bar{l}$,  $\mathcal{H}$ relies on $\bar{l}^{+}$ through the propagator
\begin{equation}
\frac{N(k)}{2\bar{l}^{+}\bar{l}^{-}+(\bar{l}_{\perp})^{2}+i\epsilon}.
\end{equation}
and the two vertexes
\begin{equation}
\frac{1}{-\bar{l}^{+}+\ldots+q_{i}^{+}+\ldots+i\epsilon},\quad\frac{1}{\bar{l}^{+}+\ldots+q_{j}^{+}+\ldots+i\epsilon},
\end{equation}
where $q_{i}$ and $q_{j}$ represent momenta of Glauber gluons and ultrasoft particles, which are defined as flow into $\bar{l}$. We have dropped terms independent of $\bar{l}^{+}$. We take $\bar{l}^{-}>0$ as one can always choose the direction of $\bar{l}$ so that $\bar{l}^{-}>0$($\bar{l}^{-}\ne 0$ unless $\bar{l}$ is a Glauber gluon or ultrasoft particle).  We can then integrate out $\bar{l}^{+}$ by taking the residue of the pole in the upper half plane.  After this operation, we can make the substitution
\begin{equation}
\bar{l}^{+}\to q_{i}^{+}+\ldots +i\epsilon
\end{equation}
in remaining terms. That is
\begin{eqnarray}
&&\frac{1}{-\bar{l}^{+}+\ldots+q_{i}^{+}+\ldots+i\epsilon}\frac{1}{\bar{l}^{+}+\ldots+q_{j}^{+}+\ldots+i\epsilon}\frac{N(\bar{l})}{2\bar{l}^{-}\bar{l}^{+}+(\bar{l}_{\perp})^{2}+i\epsilon}
\nonumber\\
&\to& \frac{-\pi iN(\bar{l})}{\bar{l}^{-}}\frac{1}{q_{i}^{+}+\ldots+q_{j}^{+}+\ldots+i\epsilon}\frac{1}{q_{i}^{+}+\ldots+\frac{(\bar{l}_{\perp})^{2}}{2\bar{l}^{-}}+i\epsilon}.
\end{eqnarray}

We repeat the procedure and get terms shaped like
\begin{equation}
\label{fincL-}
\frac{1}{\bar{P}^{+}+\sum q_{i}^{+}+\sum\frac{(\bar{k}_{i\perp})^{2}}{2\bar{k}_{i}^{-}}+\sum\frac{(\bar{l}_{i\perp})^{2}}{2\bar{l}_{i}^{-}}+i\epsilon}
\end{equation}
on left of the final cut
\begin{equation}
\label{fincR-}
\frac{1}{\bar{P}^{+}+\sum q_{j}^{+}+\sum\frac{(\bar{k}_{j\perp})^{2}}{2\bar{k}_{j}^{-}}+\sum\frac{(\bar{l}_{j\perp})^{2}}{2\bar{l}_{j}^{-}}-i\epsilon}
\end{equation}
on right of the final cut, where $q_{i}$ ($q_{j}$) represent momenta of Glauber and soft(ultrasoft) gluons and $\bar{l}_{i}$($\bar{l}_{j}$) represent momenta of collinear particles and $\bar{k}_{i}$($\bar{k}_{j}$) represent momenta of on-shell $-$-collinear particles. The summation of $\bar{l}_{i}$($\bar{l}_{j}$) is made over states with the coordinates $\widetilde{n_{+}}\cdot x$ between two given vertexes.    The summation of $q_{i}$ ($q_{j}$) and $\bar{k}_{i}$($\bar{k}_{j}$) are made over states connecting to vertexes with the coordinates $\widetilde{n_{+}}\cdot x$ smaller than those of  $\bar{l}_{i}$($\bar{l}_{j}$).

Terms of the type (\ref{fincL+}) and (\ref{fincR+}) correspond to contributions of states with the coordinates $x^{-}$(or the coordinates $\widetilde{n_{+}}\cdot x$ as $\widetilde{n_{+}}\cdot x=\frac{1}{\sqrt{2\lambda \omega}} x^{-}(1+O(\lambda))$)  smaller than that of the hard collision in $x^{-}$-ordered perturbation theory. $-$-collinear states with the coordinates $\widetilde{n_{+}}\cdot x$  greater than that of the hard collision cancel out in $\mathcal{H}(P,\bar{P},q^{+},q^{+})$.

%

\subsection{Glauber graphs in $\mathcal{H}(P,\bar{P},q^{+},q^{-})$}
\label{Glaubergh}

According to discussions in Sec.\ref{CA+G}, Glauber gluons $A_{+G}$ should couple to other particles at vertexes with the coordinates $\widetilde{n_{+}}\cdot x$ greater than those free from $A_{+G}$(except for vertexes in the hard subgraph). The result is helpful for us to exclude graphs in which the summation over all spectator-spectator and spectator-soft Glauber exchanges involving  $A_{+G}$ is hampered by spectator-active or soft-active coherence.

\begin{figure*}
\begin{center}
\includegraphics[scale=0.5]{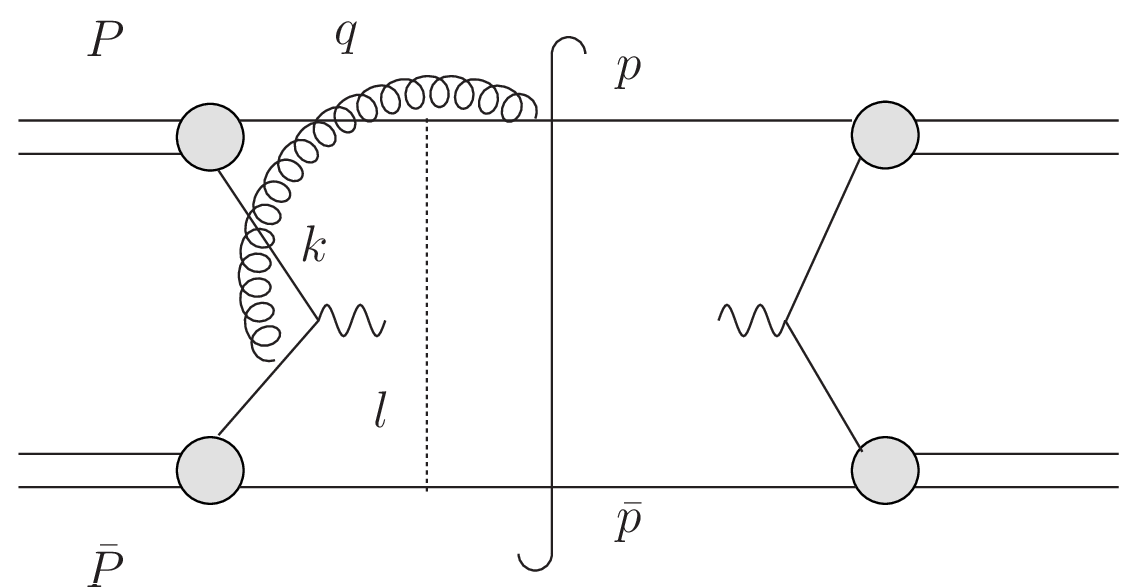}
\end{center}
\caption{An example of diagrams with couplings involving Glauber gluons couple to between couplings free from Glauber gluons and the final cut, where dot lines represent Glauber gluons and circle lines represents collinear gluons}
\label{CG+f}
\end{figure*}

Let us consider an example shown in Fig.\ref{CG+f}, in which there Glauber coupling of a plus-collinear particle before a non-Glauber coupling(not the hard vertex) of the collinear particle.   At leading order of $\lambda$ and $\eta$ we can omit the regulator terms of the form shown in (\ref{regulator}).
According to the substitution rule (\ref{app+n+}) we have
\begin{eqnarray}
Fig.\ref{CG+f}
&=&\int\frac{\ud l^{-}}{2\pi}\int\frac{\ud p_{1}^{-}}{2\pi}
\int\frac{\ud p_{2}^{-}}{2\pi}
\int\frac{\ud q^{-}}{2\pi}\int\frac{\ud k^{-}}{2\pi}
\nonumber\\
&&\frac{1}{2q^{+}q^{-}+(q_{\perp})^{2}+i\epsilon}
\frac{1}{2(P^{+}-p^{+}-q^{+})k^{-}+(P_{\perp}-p_{\perp}-q_{\perp}+l_{\perp})^{2}+i\epsilon}
\nonumber\\
&&\frac{1}{2(p^{+}+q^{+})p_{2}^{-}+(p_{\perp}+q_{\perp}-l_{\perp})^{2}+i\epsilon}
\frac{1}{2(p^{+}+q^{+})p_{1}^{-}+(p_{\perp}+q_{\perp})^{2}+i\epsilon}
\nonumber\\
&&\frac{1}{P^{-}-k^{-}-p_{2}^{-}+i\epsilon}\frac{1}{p_{2}^{-}+l^{-}-p_{1}^{-}+i\epsilon}
\frac{1}{p_{1}^{-}-q^{-}+\frac{(p_{\perp})^{2}}{2p^{+}}+i\epsilon}
\nonumber\\
&&\times \text{(terms independent of $l^{-}$ and $p_{j}^{-}$ and $q^{-}$ and $k^{-}$)($j=1,2$)}.
\end{eqnarray}
after cancelation of interactions with the coordinates $\widetilde{n_{+}}\cdot x$ greater than that of the hard collision.
We integrate out $q^{-}$ and $k^{-}$ and $p_{j}^{-}$($j=1,2$) and have
\begin{eqnarray}
\label{l--}
Fig.\ref{CG+f}
&=&\int\frac{\ud l^{-}}{2\pi}
\frac{\theta(q^{+})}{2q^{+}(p_{2}^{-}+l^{-}+\frac{(p_{\perp})^{2}}{2p^{+}})+(q_{\perp})^{2}+i\epsilon}
\nonumber\\
&&\frac{\theta(P^{+}-p^{+}-q^{+})}{2(p^{+}+q^{+})p_{2}^{-}+(p_{\perp}+q_{\perp}-l_{\perp})^{2}+i\epsilon}
\nonumber\\
&&\frac{1}{2(p^{+}+q^{+})(p_{2}^{-}+l^{-})+(p_{\perp}+q_{\perp})^{2}+i\epsilon}
|_{p_{2}^{-}=P^{-}-\frac{(P_{\perp}-p_{\perp}-q_{\perp}+l_{\perp})^{2}}{2(P^{+}-p^{+}-q^{+})}}
\nonumber\\
&&\times \text{(terms independent of $l^{-}$)}.
\end{eqnarray}
We see that all poles of $l^{-}$ locate on the lower half plane  and have
\begin{eqnarray}
Fig.\ref{CG+f}&=&0.
\end{eqnarray}

For general diagrams, one may consider flow of the plus momenta of plus-collinear particles. According to discussions in Sec.\ref{x+evo+}, plus momenta should flow from these collinear particles to the hard vertex.  We consider couplings involving these collinear particles through the order of the flow of the plus momenta\footnote{According to the physical picture in \cite{CN:1965,S:1993}, such order should be definite on pinch singular surfaces.}. We can define the momenta of these particles so that plus momenta of these particles are positive. We then integrate out minus momenta of these collinear particles by taking residues of poles of the vertex  which the plus momenta of the collinear particles flow out of. If we meet Glauber couplings before  non-Glauber couplings in some diagrams then there are some Glauber gluons of which all poles of the minus momenta(no less than two) locate in the upper half plane as shown in the example Fig.\ref{CG+f} and the formula (\ref{l--})\footnote{These poles originate from propagators of plus collinear particles coupling to Glauber gluons and vertexes which plus momenta of these particles flow into.}. Hence these diagrams vanishes after cancellation of interactions with the coordinates $\widetilde{n_{+}}\cdot x$ greater than that of the hard collision.

For Glauber couplings of minus-collinear and soft particles, one may consider flow of the minus momenta of these particles. According to discussions in Sec.\ref{x+evo-S}, we can consider these couplings through the order of the flow of the minus momenta and meet the hard vertex finally.  We define the momenta of these particles so that minus momenta of these particles are positive. Be similar to the case for Glauber couplings of plus-collinear particles,  if we meet Glauber couplings before non-Glauber couplings in some diagrams then  all poles of the plus momenta(no less than two) of some Glauber gluons locate in the upper half plane. These diagrams do not contribute to $\mathcal{H}(P,\bar{P},q^{+},q^{-})$.

In summary, spectator-active and soft-active coherence with the coordinates $\widetilde{n_{+}}\cdot x$ greater than those of Glauber interactions cancel out in $\mathcal{H}(P,\bar{P},q^{+},q^{-})$. As a result, the summation over final states in $\mathcal{H}(P,\bar{P},q^{+},q^{-})$ is inclusive enough for spectator-spectator and spectator-soft Glauber interactions involving $A_{+G}$ even if there are detected final states. This is crucial in cancellation of Glauber interactions.

\subsection{Cancellation of spectator-spectator and spectator-soft Glauber subgraphs}
\label{canGgraph}

After excluding effects of spectator-active and soft-active coherence, the Glauber subgraphs of $A_{+G}$ factorize from other parts of $\mathcal{H}(P,\bar{P},q^{+},q^{-})$. We can then consider the subgraphs separately.

\begin{figure*}
\begin{center}
\includegraphics[scale=0.4]{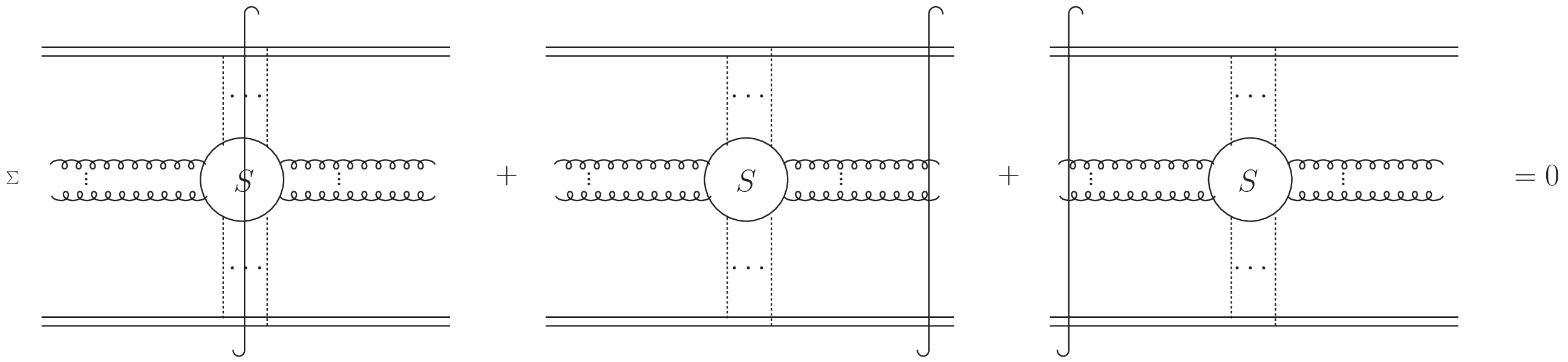}
\end{center}
\caption{ An example of cancellation of spectator-spectator and spectator-soft Glauber subgraphs.}
\label{fincsspt}
\end{figure*}

An example of cancellation of spectator-soft and spectator-spectator Glauber exchange involving $A_{+G}$ is shown in Fig.\ref{fincsspt}.
The cancellation in Fig.\ref{fincsspt} is the direct result of optical theorem if initial particles are on-shell.  Considering that momenta square of these particles are of order $Q^{2}\lambda^{2}$, the graph in Fig.\ref{fincsspt} should vanishes at leading order of $\lambda$ .

\begin{figure*}
\begin{tabular}{cc}
\includegraphics[scale=0.4]{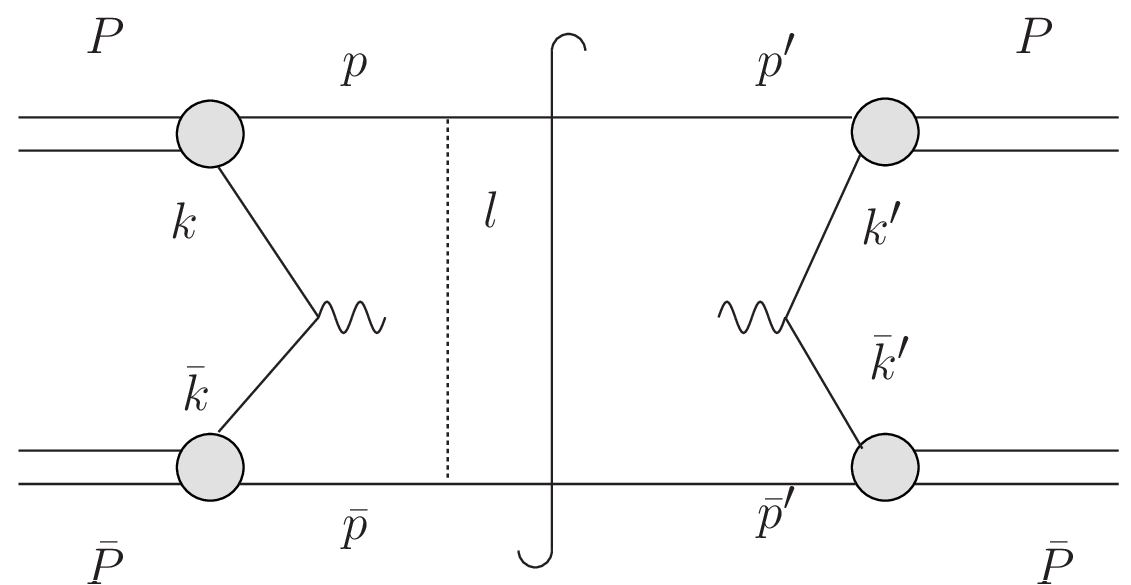}
&
\includegraphics[scale=0.4]{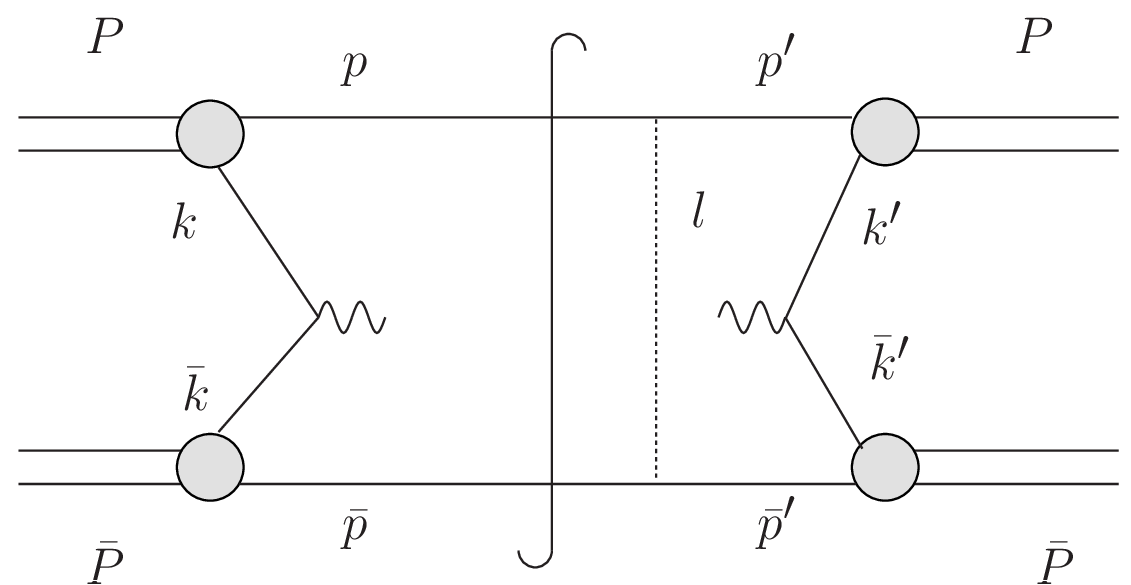}
\\
(a)&(b)
\end{tabular}
\caption{ An example of summation over spectator-spectator elastic scattering.}
\label{cangraphss}
\end{figure*}

To confirm the cancellation of spectator-spectator and spectator-soft Glauber gluons no matter the initial spectators of the Glauber exchange subprocess  are on-shell or not, let us first consider an example shown in  Fig.\ref{cangraphss}.
We have
\begin{eqnarray}
Fig.\ref{cangraphss}&=&\int\frac{\ud l^{+}}{2\pi}\int\frac{\ud l^{-}}{2\pi}\int\frac{\ud p^{-}}{2\pi}\int\frac{\ud \bar{p}^{+}}{2\pi}
\int\frac{\ud p^{\prime -}}{2\pi}\int\frac{\ud \bar{p}^{\prime +}}{2\pi}
\nonumber\\
&&
\frac{i}{P^{-}-k^{-}-p^{-}+i\epsilon}\frac{i}{\bar{P}^{+}-\bar{k}^{+}-\bar{p}^{+}+i\epsilon}
\nonumber\\
&&
\left(
\frac{i\delta(p^{\prime -}+\frac{(p_{\perp}^{\prime})^{2}}{2p^{\prime +}})}{p^{-}+\frac{(p_{\perp})^{2}}{2p^{+}}+i\epsilon}
\frac{i\delta(\bar{p}^{\prime +}+\frac{(\bar{p}_{\perp}^{\prime})^{2}}{2\bar{p}^{\prime -}})}
{\bar{p}^{+}+\frac{(\bar{p}_{\perp})^{2}}{2\bar{p}^{-}}+i\epsilon}
\frac{i}{p^{-}-l^{-}-p^{\prime -}+i\epsilon}\right.
\nonumber\\
&&
\frac{i}{\bar{p}^{+}+l^{+}-\bar{p}^{\prime +}+i\epsilon}
\frac{-i}{l_{\perp}^{2}}
+\frac{i}{l_{\perp}^{2}}
\frac{-i}{\bar{p}^{\prime +}-\bar{p}^{+}-l^{+}-i\epsilon}
\nonumber\\
&&
\left.
\frac{-i}{p^{\prime -}+l^{-}-p^{-}-i\epsilon}
\frac{-i\delta(\bar{p}^{+}+\frac{(\bar{p}_{\perp})^{2}}{2\bar{p}^{-}})}
{\bar{p}^{\prime +}+\frac{(\bar{p}_{\perp}^{\prime})^{2}}{2\bar{p}^{\prime -}}-i\epsilon}
\frac{-i\delta(p^{-}+\frac{(p_{\perp})^{2}}{2p^{+}})}{p^{\prime -}+\frac{(p_{\perp}^{\prime})^{2}}{2p^{\prime +}}+i\epsilon}
\right)
\nonumber\\
&&
\frac{-i}{P^{-}-k^{\prime -}-p^{\prime -}-i\epsilon}\frac{-i}{\bar{P}^{+}-\bar{k}^{\prime +}-\bar{p}^{\prime +}-i\epsilon}
\nonumber\\
&&\times \text{(terms independent of $p^{-}$ and $p^{\prime -}$ and $l^{-}$ and  $\bar{p}^{+}$ and $\bar{p}^{\prime +}$ and $l^{+}$)},
\end{eqnarray}
where the $i\epsilon$ term in the propagator of $l$ has been dropped as Glauber gluons are off-shell. After integrating out $l^{+}$ and $l^{-}$, we have
\begin{eqnarray}
\label{s-reli}
Fig.\ref{cangraphss}&=&\int\frac{\ud p^{-}}{2\pi}\int\frac{\ud \bar{p}^{+}}{2\pi}
\int\frac{\ud p^{\prime -}}{2\pi}\int\frac{\ud \bar{p}^{\prime +}}{2\pi}
\nonumber\\
&&
\frac{i}{P^{-}-k^{-}-p^{-}+i\epsilon}\frac{i}{\bar{P}^{+}-\bar{k}^{+}-\bar{p}^{+}+i\epsilon}
\nonumber\\
&&
\left(
\frac{i\delta(p^{\prime -}+\frac{(p_{\perp}^{\prime})^{2}}{2p^{\prime +}})}{p^{-}+\frac{(p_{\perp})^{2}}{2p^{+}}+i\epsilon}
\frac{i\delta(\bar{p}^{\prime +}+\frac{(\bar{p}_{\perp}^{\prime})^{2}}{2\bar{p}^{\prime -}})}
{\bar{p}^{+}+\frac{(\bar{p}_{\perp})^{2}}{2\bar{p}^{-}}+i\epsilon}
\frac{-i}{l_{\perp}^{2}}
\right.
\nonumber\\
&&
\left.
+\frac{i}{l_{\perp}^{2}}
\frac{-i\delta(\bar{p}^{+}+\frac{(\bar{p}_{\perp})^{2}}{2\bar{p}^{-}})}
{\bar{p}^{\prime +}+\frac{(\bar{p}_{\perp}^{\prime})^{2}}{2\bar{p}^{\prime -}}-i\epsilon}
\frac{-i\delta(p^{-}+\frac{(p_{\perp})^{2}}{2p^{+}})}{p^{\prime -}+\frac{(p_{\perp}^{\prime})^{2}}{2p^{\prime +}}+i\epsilon}
\right)
\nonumber\\
&&
\frac{-i}{P^{-}-k^{\prime -}-p^{\prime -}-i\epsilon}\frac{-i}{\bar{P}^{+}-\bar{k}^{\prime +}-\bar{p}^{\prime +}-i\epsilon}
\nonumber\\
&&\times \text{(terms independent of $p^{-}$ and $p^{\prime -}$ and  $\bar{p}^{+}$ and $\bar{p}^{\prime +}$)}.
\end{eqnarray}
We then integrate out $p^{-}$ and $p^{\prime -}$ and  $\bar{p}^{+}$ and $\bar{p}^{\prime +}$ and have
\begin{eqnarray}
Fig.\ref{cangraphss}&=&0.
\end{eqnarray}
That is, Glauber exchanges cancel out in Fig.\ref{cangraphss} no matter the spectators $p$ and $\bar{p}$ and $p^{\prime}$ and $\bar{p}^{\prime}$ are on-shell or not.

For general cases, one may consider Glauber couplings according to the $\widetilde{n_{+}}\cdot x$-order.  We take the substitution (\ref{substitutioncon+}) in couplings involving plus-collinear particles. After the substitution, minus momenta of  plus collinear particles are independent to each other. We then integrate out minus momenta of Glauber gluons coupling to plus-collinear particles by taking residues of the poles of the vertexes at which the Glauber gluons couple to plus-collinear particles. These vertexes look like
\begin{equation}
\frac{-i}{l_{i}\cdots+i\epsilon}.
\end{equation}
After these integrals, the Glauber exchange subprocess relies on minus momenta of their initial plus-collinear spectators only through their propagators or wave function . In other words,  the whole process(including the Glauber and non-Glauber subprocess) relies on the minus momenta of the spectators only through the vertex at which the spectators are produced and propagators or wave functions of the spectators.\footnote{For example, one may check $p^{-}$  terms in  (\ref{s-reli}).}   We then integrate out  minus momenta of the spectators by taking the poles of the propagators(if the spectators are off-shell) or using the on-shell condition(if the spectators are on-shell). Obviously, these two results are equivalent to each other.     For Glauber couplings of other particles, we have the similar results.  That is, cancellation of spectator-spectator and spectator-soft Glauber gluons is irrelevant to offshellness of initial particles of the elastic scattering subprocess originating form Glauber exchange at leading power of $\lambda$ and $\eta$.

Generally, one may consider the  $\widetilde{n_{+}}\cdot x$-evolution of spectators and soft particles. Such evolution is unitary as $\widetilde{n_{+}}^{\mu}$ is time like. As a result,  one has a $\widetilde{n_{+}}\cdot x$ version of optical theorem
\begin{equation}
\label{optical+}
T_{+}^{\phantom{+}\dag}T_{+}=-i(T_{+}-T_{+}^{\phantom{+}\dag}),
\end{equation}
where $T_{+}$ represents  the $\widetilde{n_{+}}\cdot x$-version of scattering matrix.  According to the relation (\ref{optical+}), Glauber exchange with the coordinates $\widetilde{n_{+}}\cdot x$ greater than those of active-spectator and active-soft coherence-summation of the Glauber exchange is not affected by the coherence-cancel out at leading order of $\lambda$. The optical theorem (\ref{optical+}) is the direct result of the unitarity of the $\widetilde{n_{+}}\cdot x$-evolution operator $U_{+G}(\infty,-\infty)$ in (\ref{canHA+G}). Cancellation in Fig.\ref{fincsspt} is a special case of (\ref{optical+}). That is
\begin{eqnarray}
&&U_{+G}^{\phantom{\dag}}(\infty,-\infty)U_{+G}(\infty,-\infty)=1
\nonumber\\
&\Rightarrow& T_{+}^{\phantom{+}\dag}T_{+}=-i(T_{+}-T_{+}^{\phantom{+}\dag}).
\end{eqnarray}
Hence graphic cancellation of spectator-spectator and spectator-soft Glauber exchange involving $A_{+G}$ based on the relation (\ref{optical+})  is equivalent to our operator level method in Sec.\ref{canA+G}.

After cancellation of Glauber gluons $A_{+G}$, one may consider an effective theory free from $A_{+G}$. Calculations of $\mathcal{H}(P,\bar{P},q^{+},q^{-})$ in this effective theory is equivalent to those in the theory (\ref{effective action}). One can then consider the $\widetilde{n_{-}}\cdot x$-evolution and  repeat the procedure of  $A_{+G}$ to see the cancellation of $A_{-G}$.  The cancellation originates from the $\widetilde{n_{-}}\cdot x$ version of optical theorem, which is equivalent to our operator level method.

\section{Conclusions and Discussions}
\label{conc}

We discuss Glauber gluon effects in hadron collisions in this paper. It is proved that effects of interactions after the hard collision cancel out in processes inclusive enough. For a time-like evolution(like $\widetilde{n_{+}}\cdot x$-evolution in Sec.\ref{x+H})  which is approximately equivalent to $x^{+}$ or $x^{-}$-evolution, we have the similar result. That is, interactions with the coordinates $\widetilde{n}\cdot x$ greater than that of the hard collision cancel out in processes inclusive enough.
After the cancellation, spectator-active coherence no longer disturb us and we prove the cancellation of spectator-spectator and spectator-soft Glauber exchanges for processes inclusive enough.    We also present the proof of eikonalization of  active-spectator and active-active and active-soft type  Glauber exchanges.  According to our discussions, these Glauber gluons should be viewed as zero-bin of collinear or soft gluons and absorbed into directions of  collinear and soft Wilson lines in hard vertex. Graphic cancellation of spectator-spectator and spectator-soft Glauber exchanges are also discussed here to show how such graphic cancellation is related to operator level skills in this paper.

Exactly, cancellation of interactions with the coordinates $\widetilde{n_{+}}\cdot x$ greater than that of the hard collision is equivalent to cancellation of final states interactions as $\widetilde{n_{+}}$ is time-like. On the other hand, $\widetilde{n_{+}}$ is approximately light-like and $\widetilde{n_{+}}\cdot x$-evolution of collinear and soft particles are approximately equivalent to light cone evolution of thee particles at mass center frame of initial hadrons as discussed in Sec.\ref{x+evo+} and Sec.\ref{x+evo-S}. This explains why proofs based on the time like evolution in this paper give the same conclusion as those based on light cone evolution in \cite{CSS:1988}.


According to our discussions, eikonal approximation is crucial in definition of collinear and soft modes at loop level.  At tree level, collinear and soft modes are characterized by their momenta. Considering that one usually run over all momenta region in practical loop integral, subtraction scheme to avoid double counting in loop level definition of different modes is necessary. Although details of such subtraction scheme is not concerned here, one loop level definitions of different modes depend on the manner they couple to tree level modes. Especially, eikonalized parts of couplings between (tree level) modes and other particles should be absorbed into those between collinear modes and those between collinear and soft or ultrasoft modes. Similarly, definitions of different modes at higher loop level depend on the manner they couple to lower loop level modes.   According to these definitions, active-spectator and active-soft exchanged gluons should be absorbed into definition of loop level collinear and soft gluons as discussed in Sec.\ref{AG},  although momenta of these gluons may be Glauber type.

Spectator-active interactions with light cone coordinates greater than  those of Glauber exchanges should be treated carefully as they may obstruct the summation of spectator-soft and spectator-spectator Glauber exchanges. Fortunately, these interactions cancel out for processes considered here as discussed in Sec.\ref{CA+G}. Intuitively, spectator-soft and spectator-spectator exchange of Glauber gluons $A_{+G}$ and $A_{-G}$ should occur at the vertexes with the coordinates $x^{+}$ and $x^{-}$ equivalent to those of the hard collision according to the classical trajectories of collinear particles\footnote{Pinch singular surfaces are related to classical trajectories as discussed in \cite{CN:1965}.} and the locality  of propagators of the Glauber gluons $A_{+G}$ and $A_{-G}$ in the $x^{+}$ and $x^{-}$ directions. Therefore one should not be surprised to see the cancellation of spectator interactions with the coordinates $\widetilde{n_{+}}\cdot x$($\widetilde{n_{-}}\cdot x$) greater than those of spectator-soft and spectator-spectator exchanges of Glauber gluons $A_{+G}$($A_{-G}$).


For electromagnetic processes like the Drell-Yan process, the result is stronger.  In these processes, there are not QCD interactions between the lepton pair $l^{+}l^{-}$ and undetected states $X$. Thus  QCD interactions between final states do not change total momentum of the lepton pair at the lowest order of electromagnetic interactions.   It is reasonable to believe that effects of interactions after the hard collision do cancel out in these processes even if one does not make the integral over $q_{\perp}$ in (\ref{intg}). This may be why Glauber couplings do not affect the transverse-momenta-dependent factorization in Drell-Yan process.

Graphic  cancellation of spectator-spectator and spectator-soft Glauber exchanges are discussed here to explain  the graphic  correspondence of the operator level proofs in this paper. Although these discussion should not be viewed as strict proofs of graphic cancellation of Glauber gluons, they are helpful for understanding  graphic aspects of key points between operator level skills in paper paper, they are(1)the unitarity of QCD evolution;(2)the cancellation of final sates interactions and interactions with light cone coordinates greater than that of the hard collision;(3)effects of the spectator-active and soft-active coherence on  spectator-spectator and soft-spectator Glauber exchanges;(4)the unitary of the evolution induced by the Glauber exchanges.

\section*{Acknowledgments}
 The work of G. L. Zhou is supported by The National Nature Science Foundation of China under Grant No. 11805151 and The Scientific Research Foundation for the Doctoral  Program of Xi'an University of Science and Technology under Grant No. 6310116055 and The Scientific Fostering Foundation of Xi'an University of Science and Technology under Grant No. 201709. The work of Z. X. Yan is supported by  The Department of Shanxi Province Natural Science Foundation of China under Grant No.2015JM1027. The work of F. Li is supported by China Postdoctoral Foundation under    Grant No. 2015M581824, 20161001  and International Postdoctoral Exchange Fellowship Program between JUELICH and OCPC.

\bibliography{scetglauber}

\end{document}